\documentclass[aps,pre,preprint,12pt]{revtex4}%
\usepackage{amsmath}
\usepackage{graphicx}
\usepackage{ifthen}
\usepackage{amsfonts}
\usepackage{amssymb}%
\def\cA{{\mathcal{A}}}

\def\bR{{\mathbb{R}}}

\def\vx{{\bf x}}

\def\be{\begin{equation}}
\def\ee{\end{equation}}
\newtheorem{theorem}{Theorem}
\newtheorem{corollary}[theorem]{Corollary}

\newtheorem{proposition}[theorem]{Proposition}
\newtheorem{lemma}[theorem]{Lemma}
\newtheorem{remark}[theorem]{Remark}
\numberwithin{equation}{section} \numberwithin{theorem}{section}
\def\qed{\hfill $\Box$\vskip 0.15in}

\begin{document}
\title{Predictability of extreme events in a branching diffusion model}
\author{A. Gabrielov} \email{agabriel@math.purdue.edu}
\affiliation{Departments of Mathematics and Earth and Atmospheric Sciences, 
Purdue University, West Lafayette, IN, 47907-1395}
\author{V. Keilis-Borok} \email{vkb@ess.ucla.edu}
\affiliation{Institute of Geophysics and Planetary Physics and Department of Earth\\ and Space Sciences,
University of California Los Angeles, 3845 Slichter Hall, Los Angeles, CA 90095-1567} 
\author{S. Olsen} \email{olsens3@unr.nevada.edu}
\author{I. Zaliapin} \email{zal@unr.edu}
\affiliation{Department of Mathematics and Statistics,
University of Nevada, Reno, NV 89557-0084.}
\date{\today}

\begin{abstract}
We propose a framework for studying predictability of extreme events in complex systems. 
Major conceptual elements --- hierarchical structure, spatial dynamics, and external driving ---
are combined in a classical branching diffusion with immigration.
New elements --- {\it observation space} and observed {\it events} --- are introduced
in order to formulate a prediction problem patterned after the geophysical and environmental 
applications. 
The problem consists of estimating the likelihood of occurrence of an extreme event given the 
observations of smaller events while the complete internal dynamics of the system 
is unknown.
We look for {\it premonitory patterns} that emerge as an extreme event approaches; 
those patterns are deviations from the long-term system's averages.
We have found a {\it single control parameter} that governs multiple 
spatio-temporal premonitory patterns.
For that purpose, we derive 
i) complete analytic description of time- and space-dependent size distribution of 
particles generated by a single immigrant; 
ii) the steady-state moments that correspond to multiple immigrants; and 
iii) size- and space-based asymptotic for the particle size distribution.  
Our results suggest a mechanism for universal premonitory patterns and provide
a natural framework for their theoretical and empirical study.

PACS numbers: 89.75.Hc, 89.75.-k, 91.30.pd, 02.50.-r, 91.62.Ty, 64.60.Ht
\end{abstract}

\maketitle

\section{Introduction}
\label{intro}
{\it Extreme events} are a most important yet least understood feature of 
natural and socioeconomic complex systems. 
In different contexts these events are also called critical transitions, 
disasters, catastrophes, or crises. 
Among examples are destructive earthquakes, El-Ni\~nos, heat waves, 
electric power blackouts, economic recessions, stock-market crashes, 
pandemics, armed conflicts, and terrorism surges. 
Extreme events are rare, but consequential: they inflict a lion's share of 
the damage to population, economy, and environment.
The present study is focused on predicting individual extreme events. 
That problem is pivotal both for fundamental understanding of complex 
systems and for disaster preparedness 
(see {\it e.g.}, \cite{KBS03,Sor04,AJK05}).

Our {\it approach to prediction} is complementary to more traditional 
and well-developed ones, which include 
classical Kolmogoroff-Wiener extrapolation of time series \cite{Kol41, Wie49}, 
linear (Kalman-Bucy) \cite{KB61} and 
non-linear (Kushner-Zakai) \cite{Kush62,Zak69,Chow07} filtering,
sequential Monte-Carlo methods \cite{DFG01}, or 
the extreme-value theory \cite{EKM08}.
The need for a novel approach is dictated by a non-standard 
formulation of the prediction problem, where one is particularly interested in
the future occurrence times of rare events rather 
than the complete unobserved state of the system in continuous time.
We notice, accordingly, that often the easily observed extreme events 
can not be defined as the instants of threshold exceedance by the observed 
physical or economical fields, like air temperature or asset price.
A paradigmatic example is an earthquake initiation time, which is determined by 
complex interplay of stress and strength fields in the heterogeneous Earth 
lithosphere. 
The physical theory for spatio-temporal evolution of these fields is still 
in its infancy, their values can hardly be measured with the existing instruments,
or predicted using the available statistical methods.
At the same time, earthquakes are readily defined, measured, and studied.

Prediction here is based on analysis of observable permanent background activity 
of the complex system. 
We look for {\it premonitory patterns}, {\it i.e.}, particular deviations from long-term 
averages that emerge more frequently as an extreme event approaches. 
These patterns might be either {\it perpetrators} contributing to triggering an 
extreme event, or {\it witnesses} merely signaling that the system became unstable, 
ripe for a disaster. 
An example of a witness is proverbial ``straws in the wind" preceding a hurricane.

The following types of premonitory patterns have been established by exploratory 
data analysis and numerical modeling: 
(i) increase of background activity; 
(ii) deviations from self-similarity: change of the size distribution of events 
in favor of relatively strong yet sub-extreme events; 
(iii) increase of event's clustering; and 
(iv) emergence of long-range correlations. 
Solid empirical evidence for existence of these patterns in seismology and other 
forms of multiple fracturing has been accumulated since the 1970s  
\cite{KBS03,KKR80,KB94,KB96,KK90,KS99,PCS94,SSS99,Vor99,ALP82,Aki85,PA95,Rom93,Mog81,Hab81,Smi81,
YK92,RTK00,KB02,Tur91,Lock93,MDR90,RKB97,Jor06,JS99,ZHK01,EB97b,ABR04}. 
Importantly, these patterns are universal, common for complex systems
of distinctly different origin.
Similar premonitory patterns have been observed in
socio-economic systems \cite{KSSM00,KSA05}, 
dynamic clustering in elastic billiards \cite{GKSZ08}, 
hydrodynamics, and hierarchical models of extreme event development 
\cite{NS90,NS94,NTG95,BSL97,GZKN00,ZKG03}.
We propose here a general mechanism that reproduces these
universal premonitory patterns.

We focus in particular on {\it premonitory deviations from self-similarity}.
Self-similarity is one of the most prominent features of complex systems.
A canonical example is a power-law (self-similar) distribution of system's 
observables, whose remarkable feature is inevitability of extremely 
large events that dwarf numerous smaller events.
Power-law distribution is well known under different names in such diverse phenomena as 
inertial-range self-similarity in turbulence (Kolmogorov-Obukhov laws) 
\cite{Kol41a,Kol41b,Obu41,Frisch,JCM90},
energy released in an earthquake (Gutenberg-Richter law) \cite{GR41,GR44,BZ03},
word usage frequency in a language (Zipf law) \cite{Zipf}, 
allocation of wealth in a society (Pareto law) \cite{Pareto,KBL+06},
war casualties (Richardson law) \cite{Rich},
number of papers published by a given scientist (Lotka law) \cite{Lotka}, 
mass of a landslide \cite{MTG+04,BGR09}, 
stock price returns \cite{BT67,PS08,GGP+03}, 
number of species per genus \cite{Bur90}, 
and many other \cite{New97,New05,AB02,Mandelbrot,Turcotte}.
An important paradigm of {\it self-organized criticality} \cite{BTW88,Tur99} 
that is demonstrated by sand-pile \cite{Dhar90}, forest-fire \cite{DS92}, and 
slider-block \cite{BK67,OFC92,RK93} models and their numerous ramifications has 
been introduced in order to understand dynamic processes whose only attractor 
corresponds to self-similarity (criticality) of the size distribution of
appropriately defined {\it events}.

Exact self-similarity, as well as many other universal properties, 
however is only an approximation to (or a mean-field property of) 
the observed and modeled systems; 
at each particular time moment the distribution of event sizes deviates 
from a pure power-law form.
We show in this paper how to use such deviations for understanding the 
dynamics of a complex system in general and occurrence of extreme
events in particular.

The rest of the paper is organized as follows.
We informally outline our model and the corresponding prediction problem
in Sect.~\ref{outline}.
A formal model description is given in Sect.~\ref{model}.
Section~\ref{sum} summarizes the study's results most relevant to the 
prediction problem.  
Section~\ref{results} derives the spatio-temporal model distribution
as a function of the control parameter.
Section~\ref{GR} uses these results to find spatio-temporal
deviations of the event size distribution from its mean-field form.
Results of numerical experiments are illustrated in Sect.~\ref{num}.
In Section~\ref{discussion} we further discuss the relation of our results
to prediction of extreme events.
Proofs and necessary technical information are collected in Appendices.

\section{Model outline and prediction problem}
\label{outline}
Our model combines {\it external driving} ultimately responsible for occurrence 
of events, including the extreme ones, a {\it cascade process} responsible for 
redistribution of energy (or another appropriate physical quantity such as
mass, moment, stress, {\it etc.}) within the system, and {\it spatial dynamics}.  
We first outline the process of populating a system space $\Omega$ with 
{\it particles} of discrete ranks and then proceed with definition of the 
{\it observation space} and {\it events}.
We assume that $\Omega$ is an $n$-dimensional Euclidean space.

A direct cascade (branching) within a system starts with consecutive injection
(immigration) of {\it particles} of the largest possible rank, $r_{\rm max}$, 
into the origin ${\bf 0}\in\Omega$, which we call {\it source}.
After injection, each particle diffuses freely and independently 
of the others across the space $\Omega$.
Eventually, it splits into a random number of particles of smaller
rank, $r_{\rm max}-1$, each of which continues to diffuse from the location 
of the parent and independently of the other particles.
These particles split in their turn into even smaller particles,
and so on.

At each time instant $t\ge 0$, {\it observations} can be done on a 
subspace $\mathcal{R}_t\subset\Omega$.
In this paper we assume that $\mathcal{R}_t$ is an affine subspace of
dimension $d<n$.
An observed {\it event} corresponds to an instant when a particle 
crosses the subspace of observations.
Each event is characterized by its occurrence time $t$, spatial
location ${\bf x}\in\mathcal{R}_t$ within the observation space, and
rank $r$.
Model observations at instant $t$ thus consist of a collection of events 
$\mathcal{C}_t=(t_i\le t,{\bf x}_i,r_i)$, $i\ge 1$, referred to as
{\it catalog}.
{\it Extreme event} is defined as a sufficiently large, although
not necessarily the largest, event, $r\ge r_0$,
where $r_0$ is a rank threshold.

Importantly, the location of $\mathcal{R}_t$ within $\Omega$ is
a) not known to the observer, and b) time-dependent.
One can interpret this as movement of the observation space
relative to the source, movement of the source relative to
the observation space, or combination of the two.
A principal goal of an observer is to assess the likelihood of the 
occurrence of an extreme event using the catalog $\mathcal{C}_t$.
It is readily seen that the probability of an extreme event increases 
as the observation space approaches the source and achieves its maximal 
value when the source belongs to the observation space, 
${\bf 0}\in\mathcal{R}_t$.
The distance between the observation subspace and the source thus 
becomes a natural control parameter and allows one to reduce
the prediction problem to estimating the distance to the source.
This latter problem is the focus of our study. 

\begin{figure}
\centering\includegraphics[width=.4\textwidth]{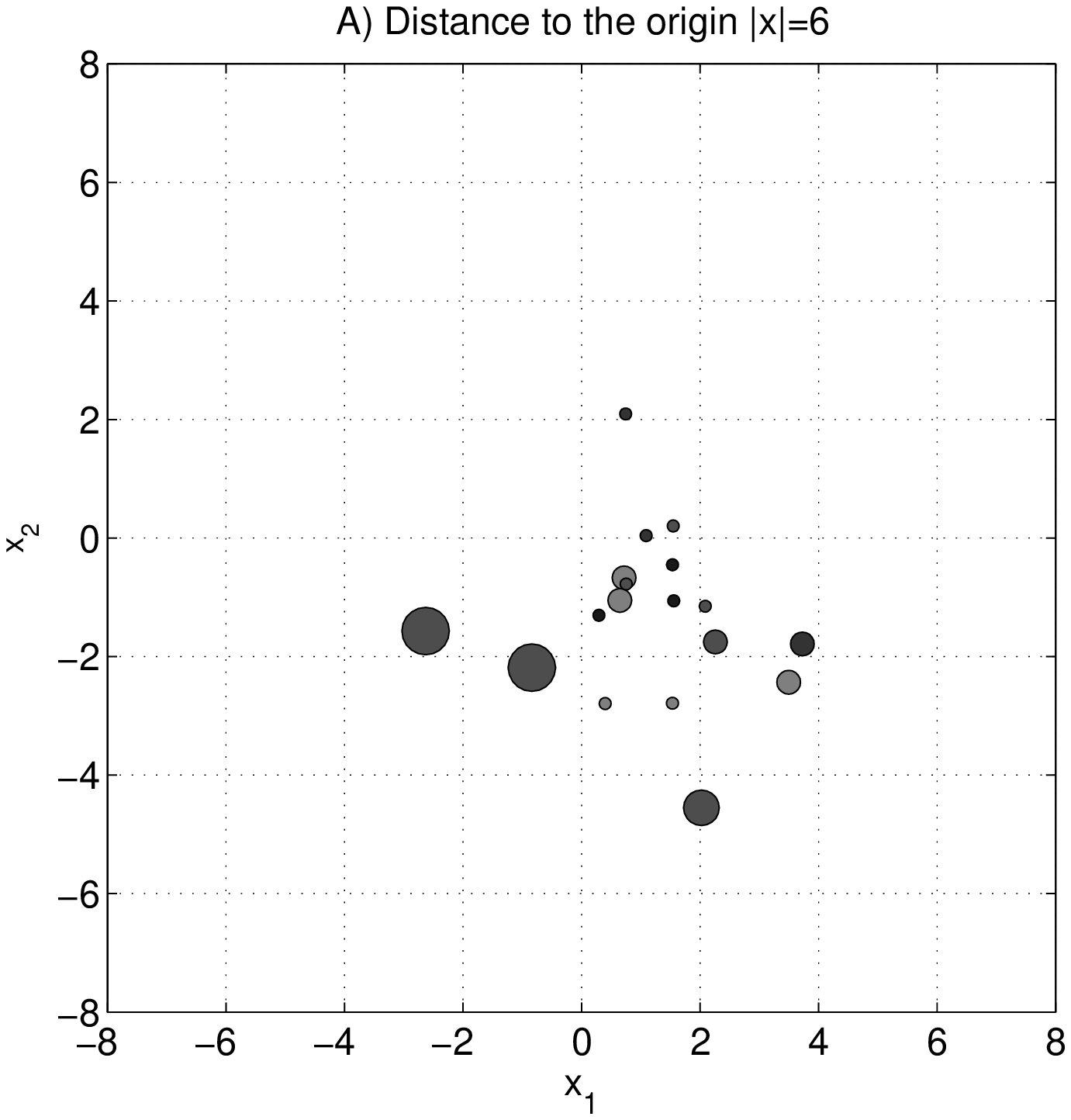}
\centering\includegraphics[width=.4\textwidth]{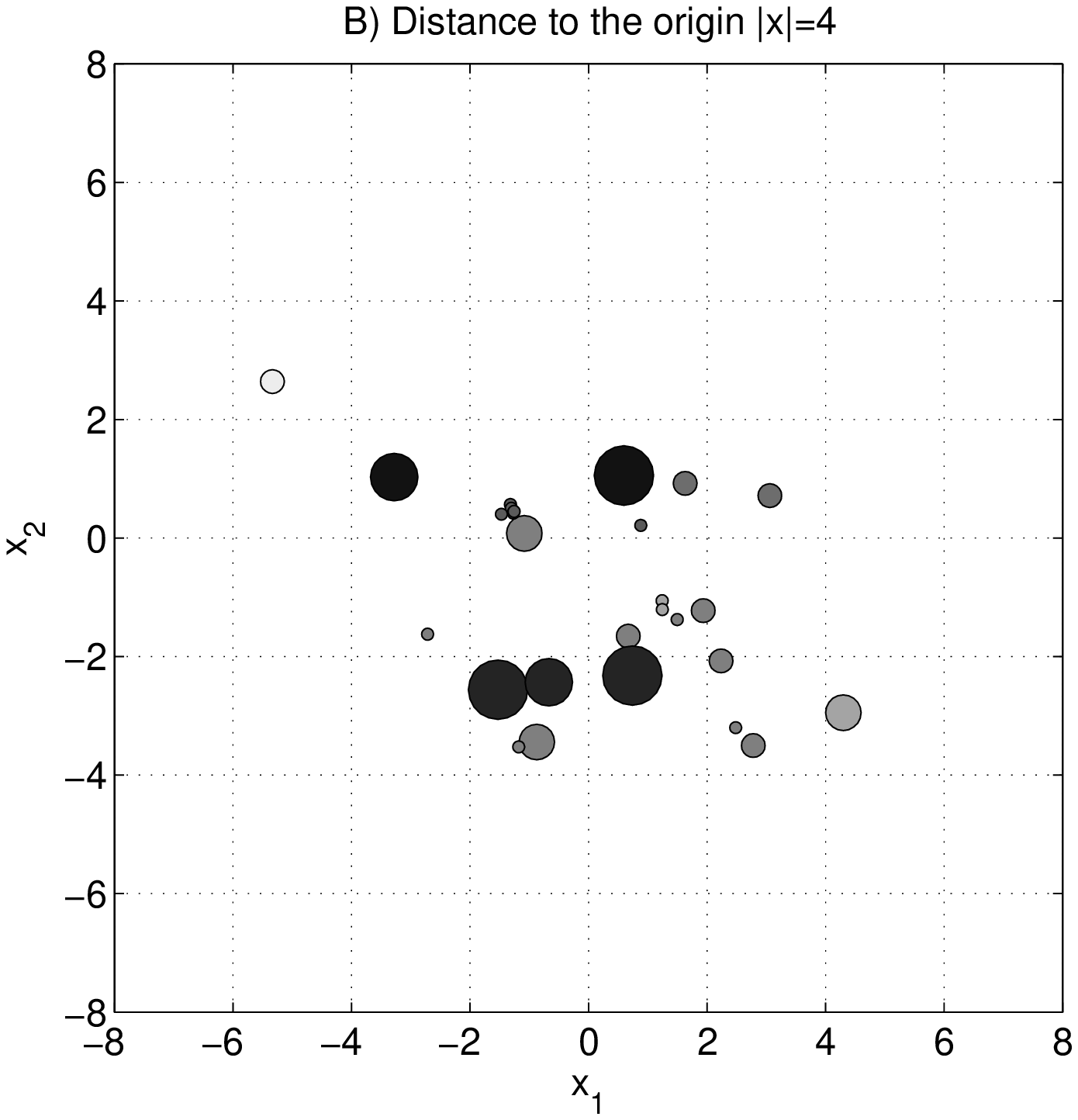}
\centering\includegraphics[width=.4\textwidth]{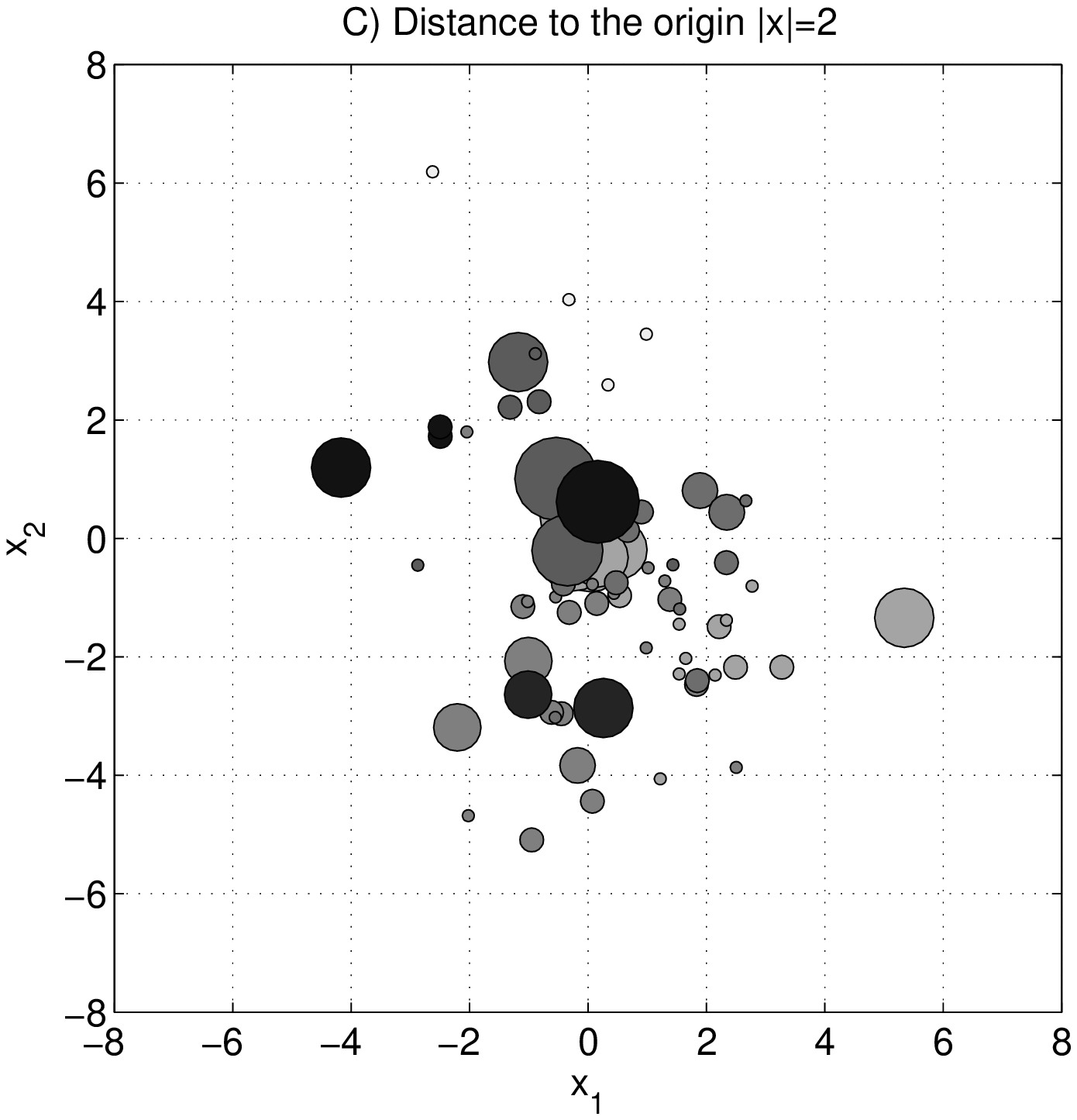}
\centering\includegraphics[width=.4\textwidth]{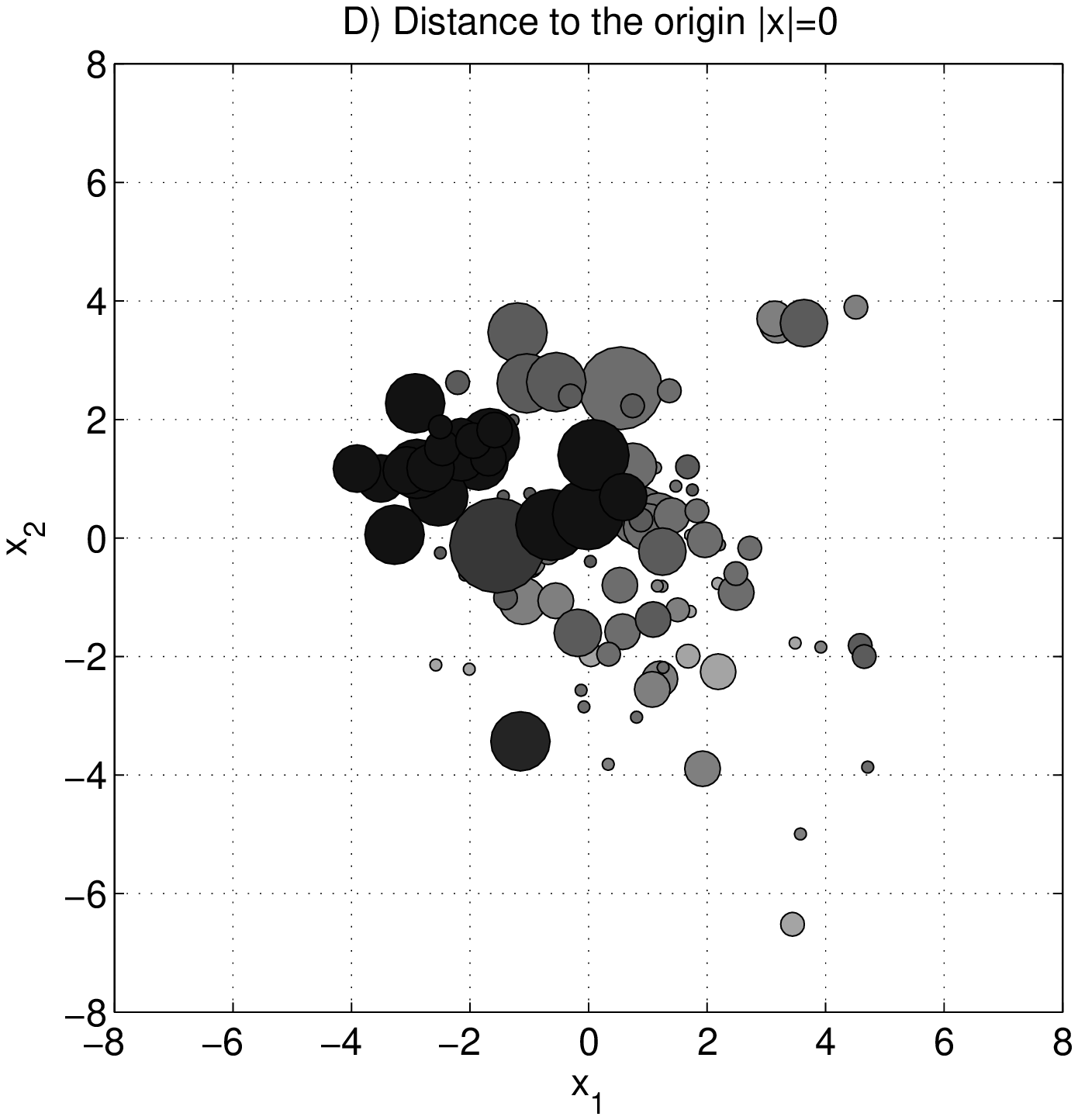}
\caption{Example of a 3D model population.
Different panels show 2D subspaces of the model 3D space
at different distances $|{\bf x}|$ to the origin. 
Model parameters are $\mu=\lambda=1$, $D=1$, $B=2$.
Circle size is proportional to the particle rank. 
Different shades correspond to populations from different 
immigrants; the descendants of earlier immigrants have lighter shade.
The clustering of particles is explained by the 
splitting histories.
Note that, as the origin approaches, the particle activity
significantly changes, indicating the increased probability
of an extreme event.
} 
\label{fig_example}
\end{figure}

As the observation subspace approaches the source, intensity of
the observed events increases, larger events become relatively more 
frequent, clustering and long-range correlations become more prominent
(see Fig.~\ref{fig_example} and Sects.~\ref{sum},\ref{discussion}).
Emergence of these patterns, each individually and all together, can be
therefore used to forecast an approach of a large event; indeed, such a 
prediction should be understood in a statistical sense.
This study is focused on quantitative description of two of these
patterns, intensity increase and deviations from self-similarity,
for a classical branching process formally introduced in the next section. 

We emphasize that the location and dynamics of the observation 
space $\mathcal{R}_t$ within $\Omega$ depend on details of a particular 
system of interest and may be hard to estimate or model.
An important result of this paper is that 
(i) the information about this unknown dynamics can be summarized 
by a scalar value of the control parameter (distance 
between the observational subspace and the origin); and 
(ii) knowledge of the control parameter is sufficient to solve 
the prediction problem.

Finally, it is important to mention that we do not use direct cascade
as a dynamical model of event formation, which would imply
that large events {\it cause} smaller ones.
We merely use this analytically tractable approach to create a {\it hierarchical 
network} of spatially distributed particles.
A dynamic interpretation of the latter will depend on a concrete application,
and may include inverse cascading or other physically relevant processes.

\section{Model formulation}
\label{model}
We consider an age-dependent multi-type branching diffusion process with
immigration in $\mathbb{R}^n$.
The system consists of particles, each of which belongs to a   
{\it generation} $k=0,1,\dots$.
Particles of zero generation (the largest ones) appear in a system as a result of 
external driving (forcing); we will refer to them as {\it immigrants}. 
Particles of any other generation $k>0$ are produced as a result of 
splitting of particles of generation $k-1$. 
Immigrants ($k=0$) are born at the origin 
${\bf x}:=(x_1,\dots,x_n) = {\bf 0}$ at 
a constant rate $\mu$;
that is, the probability for a new immigrant to appear
within the time interval of length $\Delta t$ is 
$\mu\Delta t+o(\Delta t)$ as $\Delta t\to 0$. 
Accordingly,
the birth instants form a homogeneous Poisson process 
with intensity $\mu$.
Each particle lives for some random time $\tau$ and then transforms (splits) into a 
random number $\beta$ of particles of the next generation.
The probability laws of the lifetime $\tau$ and branching $\beta$  
are generation-, time-, and space-independent.
We assume that new particles are born at the location of their 
parent at the moment of splitting.

The particle lifetime has an exponential distribution:
\be
G(t) := \mathsf{P}\{\tau<t\} = 1 - e^{-\lambda\,t},~\lambda>0.
\label{lifetime}
\ee
The conditional probability that a particle transforms into $k\ge 0$ 
new particles (0 means that it disappears) given that the transformation 
took place is denoted by $p_k$.  
The probability generating function (pgf) for the number $\beta$ of
new particles is thus
\be
\label{branching_pgf}
h(s) = \sum_k p_k\,s^k.
\ee
The expected number of offsprings (also called the {\it branching number}) 
is $B:=E(\beta)=h'(1)$ (see {\it e.g.}, \cite{AN04}, Chapter 1).

Each particle diffuses in $\mathbb{R}^n$ independently of other particles. 
This means that the density $p({\bf x,y},t)$ of a particle
that was born at instant $0$ at point ${\bf y}$ solves the equation
\be
\frac{\partial p}{\partial t}
= D\left(\sum_i \frac{\partial^2}{\partial x_i^2}\right)p
\equiv D\bigtriangleup_{\vx} p
\label{diffusion}
\ee
with the initial condition $p({\bf x,y},0) = \delta({\bf x-y})$.
The solution of (\ref{diffusion}) is given by \cite{Eva98}
\be
p({\bf x,y},t)= \left(4\,\pi\,D\,t\right)^{-n/2}
\exp\left\{-\frac{|{\bf x-y}|^2}{4\,D\,t}\right\},
\quad |{\bf x}|^2 = \sum_i x_i^2.
\label{sol}
\ee 
Accordingly, the density of each particle, given that it is alive at the instant $t$, 
is $\phi({\bf x},t):=p({\bf x},{\bf 0},t)$.
Naturally, the positions of the particles produced by the same
immigrant are correlated.
This can be reflected by the joint distribution of pairs, triplets, {\it etc}. 

The model is specified by the following parameters: immigration intensity $\mu>0$, 
branching intensity $\lambda>0$, diffusion constant $D>0$, and branching distribution 
$\{p_k\}$, which will be often represented by its pgf $h(z)$ or simply by the
branching number $B$.
An appropriate choice of the temporal and spatial scales
allows one to assume $\mu=1$ and $D=1$. 

It is convenient to introduce particle {\it rank} $r:=r_{\rm max}-k$ for an
arbitrary integer $r_{\rm max}$ and thus 
consider particles of ranks $r\le r_{\rm max}$. 
Particle rank can be considered a logarithmic measure of the size.
Similar to the analysis of the real-world systems, we sometime only
consider particles of the first several generations $0\le k \le r_{\rm max}-1$,
which corresponds to the largest ranks $1\le r \le r_{\rm max}$.
Figure~\ref{fig_example} illustrates the model population. 

\section{Summary of results related to prediction}
\label{sum}
We summarize here the study's findings that are most relevant to the 
prediction problem.
Recall that the prediction problem consists of assessing the likelihood
of an extreme event; the latter corresponds to an instant when a sufficiently
large particle crosses the observation space. 
The likelihood of an extreme event is thus directly related to the 
distance between the space of observations and the origin.
Accordingly, the prediction problem is reduced to the estimation of
this distance from available data.
For that, one should look for increase in the intensity of 
medium-to-large-sized events, as well as upward deviations in the 
event size distribution.  
We believe that this general idea can be useful in a wide range of models 
and observed systems, not necessarily based on a branching diffusion mechanism.
Statistical assessment of particular prediction schemes based on this
idea is left for a future study.

All statements below refer to a steady-state of the model (dynamics
after a transient).
All asymptotic statements have been confirmed numerically in finite models.

\begin{figure}
\centering\includegraphics[width=.8\textwidth]{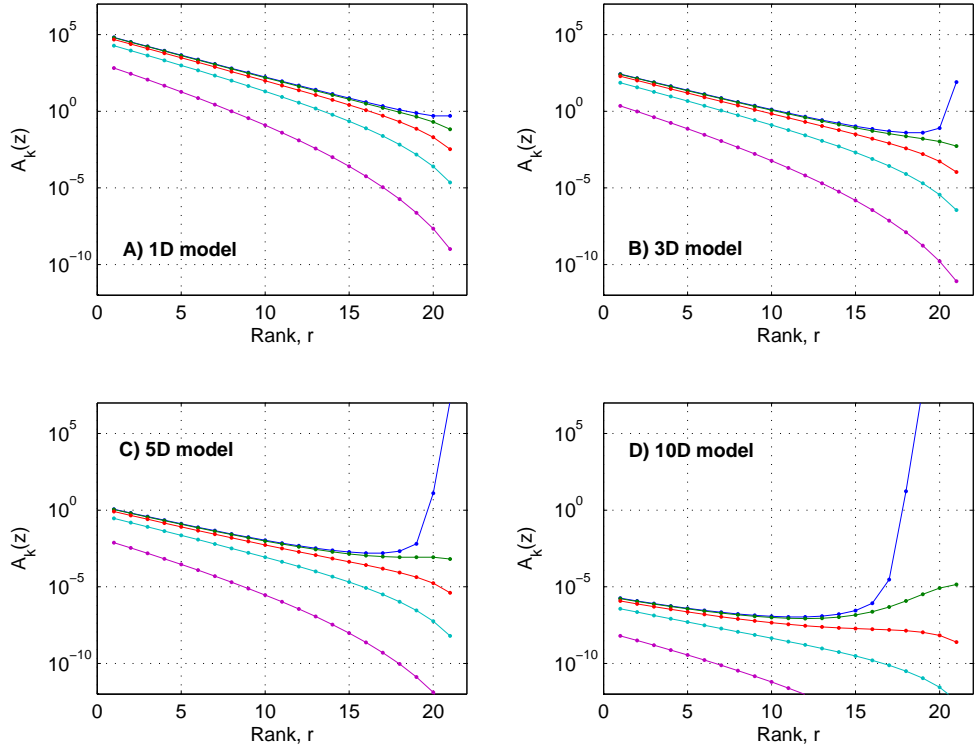}
\caption{Expected number $A_k(z)$ of generation $k$ particles 
at distance $z$ from the origin (cf. Proposition \ref{gammalim}). 
The distance $z$ is increasing
(from top to bottom line in each panel) as $z=10^{-3},2,5,10,20$.
Model dimension is $n=1$ (panel A), $n=3$ (panel B),
$n=5$ (panel C), and $n=10$ (panel D).  
Other model parameters: $\mu=\lambda=1$, $D=1$, $B=2$, $r_{\rm max}=21$. 
One can clearly see the transition from downward to upward
deviation of the rank distributions from the pure exponential form as 
we approach the origin.
Notably, the magnitude of the upward deviation close to the origin
(the upper line in all panels) strongly increases with the model 
dimension $n$.}
\label{fig_Ak}
\end{figure}

\begin{figure}
\centering\includegraphics[width=.8\textwidth]{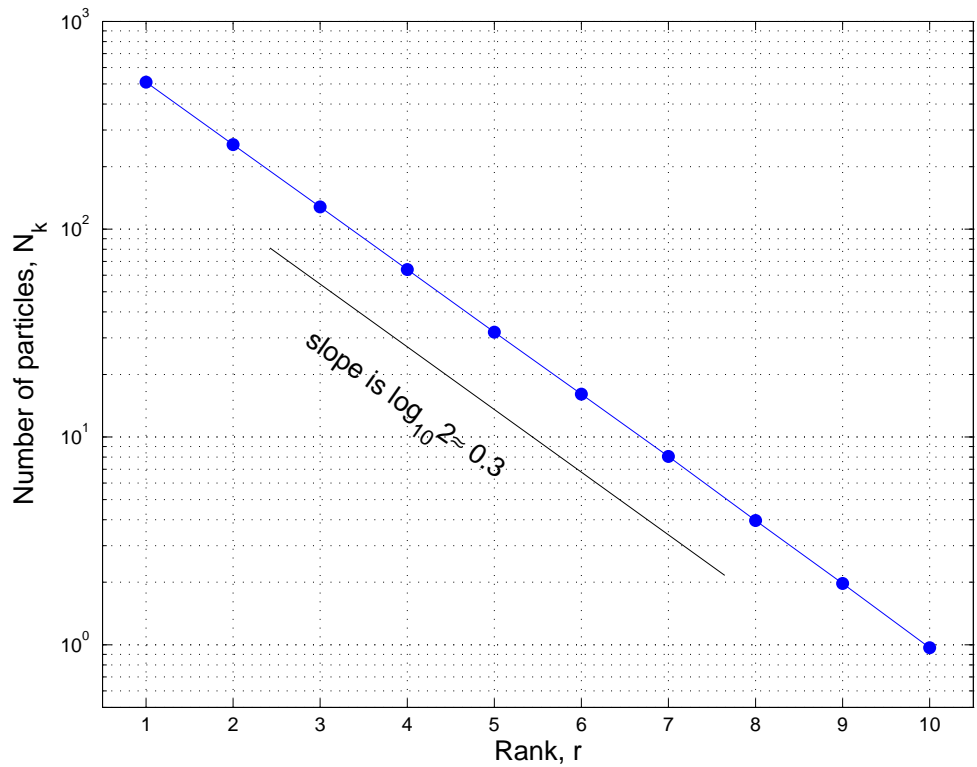}
\caption{Spatially averaged particle rank distribution at $t=30$. 
The distribution is averaged over 4000 independent realizations of a 3D model with 
parameters $\mu=\lambda=1$, $D=1$, $B=2$, $r_{\rm max}=10$. 
One can clearly see the exponential rank distribution of Eq.~\eqref{pureexp}.} 
\label{fig_GR}
\end{figure}

\begin{figure}
\centering\includegraphics[width=.8\textwidth]{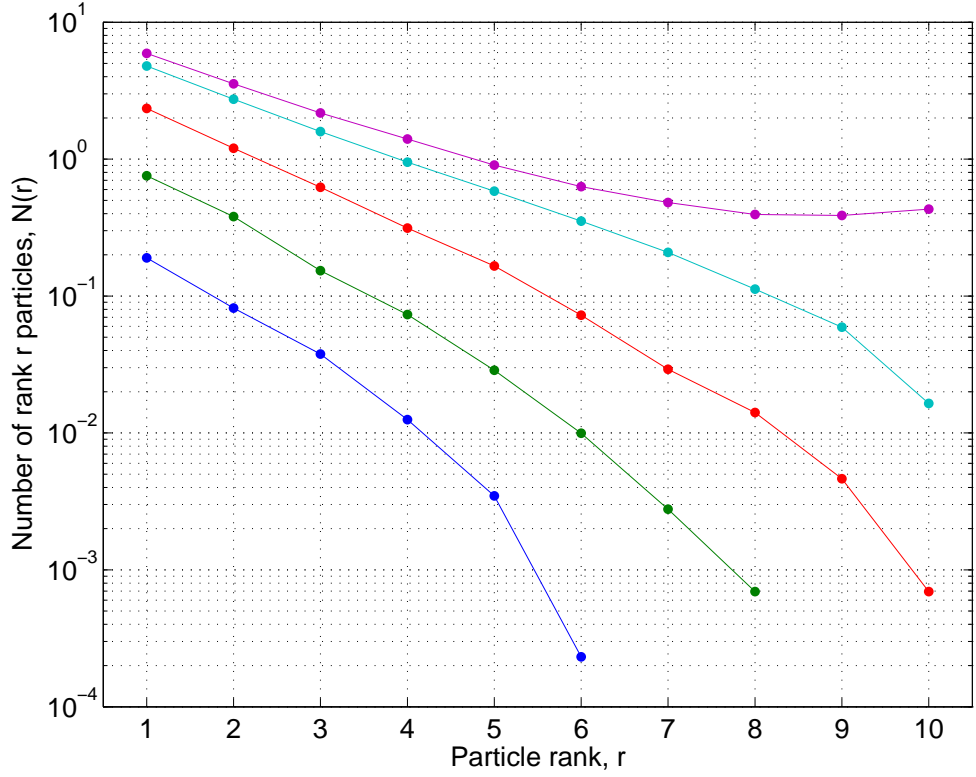}
\caption{Particle rank distribution at $t=30$ and fixed distance
$z$ from the origin (cf. Proposition \ref{gammalim}). 
The distribution is averaged over 4000 independent realizations 
of a 3D model with parameters $\mu=\lambda=1$, $D=1$, $B=2$, $r_{\rm max}=10$. 
Different lines correspond to different distances (from top
to bottom): $z=0,2,4,6,8$. 
Spatial averaging is done within spherical shells of constant 
volume $V=5$ with inner radius $z$.
One can clearly see that the rank distribution deviates from the
pure exponential form, which corresponds to a straight line in the 
semilogarithmic scale used here. 
One observes {\it downward deviations} at large distances from the 
origin, and {\it upward deviations} close to the origin.
} 
\label{fig_GR_3D}
\end{figure}

\begin{itemize}
\item[1.] {\it Meanfield self-similarity}. 
Particle ranks, averaged over time and space, have an 
exponential distribution; this is equivalent to a power-law 
distribution of particle sizes; 
see \eqref{pureexp} and Fig.~\ref{fig_GR}.
\item[2.] {\it Small-size self-similarity}.
The particle rank distribution at {\it any} spatial point is 
asymptotically exponential as rank decreases, with the exponent 
index $-B$; see \eqref{gammakinf} and 
Figs.~\ref{fig_Ak} and \ref{fig_GR_3D}.
This is equivalent to a power-law distribution of particle sizes 
with power-law index $-B$.
Furthermore, this implies that deviations from self-similarity, if any, can be 
only seen at large ranks (large particle sizes).
\item[3.] {\it Upward deviations close to the origin.}
At any point sufficiently close to the origin, the particle size 
distribution deviates from a self-similar power-law form as to have 
a larger number of medium-to-large-sized events.  
The magnitude of this deviation increases with the event size, as well 
as with dimension of the model space; see \eqref{gammaz0}
and the upper lines in Figs.~~\ref{fig_Ak} and \ref{fig_GR_3D}.
\item[4.] {\it Downward deviations away from the origin.}
At any point sufficiently far from the origin, the particle size
distribution deviates from a self-similar power-law form as to
have a smaller number of medium-to-large-sized events.
The magnitude of this deviation increases with the event size and
is independent of the model's dimension;
see \eqref{gammazinf}
and the lower lines in Figs.~\ref{fig_Ak} and \ref{fig_GR_3D}.
\item[5.] {\it Exponential decay of event intensity.}
The intensity of events of any
fixed size is exponentially decaying away from the origin; see \eqref{Az}.
\item[6.] {\it  Divergence of event intensity at the origin.}
For models with spatial dimension larger than 1, the intensity of sufficiently 
large events diverges at the origin in a power-law fashion; see \eqref{Az},\eqref{as} and
Fig.~\ref{fig_GR}(b,c,d).
\end{itemize}

\section{Model solution: Moment generating functions}
\label{results}

The model introduced in Sect.~\ref{model} is a superposition of independent 
branching processes generated by individual immigrants.
Sections~\ref{single_one} and \ref{single_two} analyze, respectively, 
the one-point and two-point moments of a particle distribution produced
by a single immigrant.
Then we expand these results to the case of multiple immigrants in 
Sect.~\ref{multiple}.

\subsection{Single immigrant: One-point properties}
\label{single_one}

\subsubsection{Moment generating functions}
Let $p_{k,i}(G, \textbf{y}, t)$ be the conditional probability that at time 
$t \geq 0$ there exist $i \geq 0$ particles of generation $k \geq 0$ within 
spatial region $G \subset \mathbb R^n$ given that at time 0 a single immigrant 
was injected at point $\textbf{y}$.  
The corresponding {\it moment generating function} is 
\begin{eqnarray}
\label{eqn:mgen2}
	M_k(G,{\bf y},t; s) = \sum_i p_{k,i}(G, {\bf y}, t)e^{si}.
\end{eqnarray}

\begin{proposition}
The moment generating functions $M_k(G, {\bf y}, t; s)$ solve the following recursive 
system of non-linear partial differential equations:
\begin{eqnarray}
\label{eqn:pdeM}
\frac{\partial}{\partial t} M_k(G, {\bf y}, t; s) = D \Delta_{{\bf y}} M_k - \lambda M_k + \lambda \, 
h(M_{k-1}), \quad k \geq 1,
\end{eqnarray}
with initial conditions 
$M_k(G, {\bf y}, 0; s) \equiv 1$, $k \geq 1$, and
\begin{eqnarray}
\label{eqn:initialM}
M_0(G, {\bf y}, t; s) = (1-P)+Pe^{s}, \quad P:=e^{-\lambda t} 
\int_G p({\bf x}, {\bf y}, t)d{\bf x}.
\end{eqnarray}
Here $h(s)$ is defined by \eqref{branching_pgf} 
and $\Delta_{{\bf y}} = \sum_i \partial^2/\partial y_i^2.$
\label{prop1}
\end{proposition}
{\bf Proof} is given in Appendix~\ref{p_prop1}.

\subsubsection{The first moment densities}
Let $\bar A_k(G, {\bf y}, t)$ be the expected number of generation-$k$ 
particles at instant $t$ within the region $G$, produced by a single immigrant injected at 
point ${\bf y}$ at time $t=0$.  
It is given by the following partial derivative (see {\it e.g.}, \cite{AN04}, Chapter 1):
\begin{eqnarray}
\label{eqn:ex1}
\bar A_k(G,{\bf y}, t) := \frac{\partial M_k(G,{\bf y}, t ; s)}{\partial s} \Big|_{s=0}.
\end{eqnarray}
Consider also the expectation density $A_k({\bf x},{\bf y},t)$ that satisfies, 
for any $G \subset \mathbb R^n$, 
\begin{eqnarray}
\label{eqn:ex2}
\bar A_k(G,{\bf y},t) = \int_G A_k({\bf x},{\bf y},t) d{\bf x}.
\end{eqnarray}

\begin{corollary}
The first moment densities $A_k({\bf x}, {\bf y}, t)$ solve the following recursive 
system of partial differential equations:
\begin{eqnarray}
\label{eqn:pdeA}
\frac{\partial A_k({\bf x}, {\bf y}, t)}{\partial t} = 
D \Delta_{{\bf x}} A_k - \lambda A_k + \lambda B A_{k-1}, \quad k \geq 1,
\end{eqnarray}
with the initial conditions $A_k({\bf x},{\bf y}, 0) \equiv 0$, $k \geq 1$, 
\begin{eqnarray}
\label{eqn:initialA}
A_0({\bf x},{\bf y}, 0) = \delta({\bf y}-{\bf x}), \quad 	
A_0({\bf x},{\bf y}, t)=e^{-\lambda t} p({\bf x},{\bf y}, t), \: t> 0.
\end{eqnarray}
The solution to this system is given by
\begin{eqnarray}
\label{eqn:solA}
A_k({\bf x},{\bf y}, t) &=& \frac{(\lambda B t)^k}{k!}A_0({\bf x},{\bf y}, t) \nonumber \\
&=& \frac{(\lambda B)^k}{k! (4 \pi D)^{n/2}}t^{k-n/2} \, 
\exp\left\{-\lambda t - \frac{|{\bf x}-{\bf y}|^2}{4Dt}\right\}.
\end{eqnarray}
\label{col11}
\end{corollary}
{\bf Proof} is given in Appendix~\ref{p_col11}.
It follows from a general result for the higher moments obtained in Appendix~\ref{p_col1}.

The system \eqref{eqn:pdeA} has a transparent intuitive meaning.
The rate of change of the expectation density $A_k({\bf x},{\bf y},t)$ is 
affected by the three processes: diffusion of the existing particles 
of generation $k$ (the first term in the rhs of \eqref{eqn:pdeA}), 
splitting of the existing particles of generation $k$ at the rate $\lambda$ 
(the second term), 
and splitting of the generation $k-1$ particles that produce on average $B$ new 
particles of generation $k$ (the third term).  

To obtain the solution for the entire population, we sum up the 
contributions from all generations:
\be
A({\bf x},{\bf 0},t)=\sum_{k=0}^\infty A_k({\bf x},{\bf 0},t)
=e^{-\lambda\,t\,(1-B)}\,p({\bf x},{\bf 0},t)=
\frac{e^{-\lambda\,t\,(1-B)}}{\left(4\,\pi\,D\,t\right)^{n/2}}
\exp\left(-\frac{|{\bf x}|^2}{4\,D\,t}\right).
\ee
This formula emphasizes the role of the branching parameter $B$:
in subcritical case, $B<1$, the population extincts exponentially; 
in supercritical case, $B>1$, the population grows exponentially;
in critical case, $B=1$, the expected number of particles remains the same
(steady state) and is given by the diffusion density $p({\bf x},{\bf 0},t)$. 

\subsection{Single immigrant: Two-point properties}     
\label{single_two}

\subsubsection{Moment generating functions}
Let $p_{k_1, k_2, i, j}(G_1, G_2, \textbf{y}, t)$ be the conditional probability that 
at instant $t \geq 0$ there exist $i \geq 0$ particles of generation $k_1 \geq 0$ within 
region $G_1 \subset \mathbb R^n$ and $j \geq 0$ particles of generation $k_2 \geq 0$ 
within region $G_2 \subset \mathbb R^n$ given that at time 0 a single immigrant was injected 
at point $\textbf{y}$.  
Assume that $G_1$ and $G_2$ do not overlap.
The corresponding moment generating function is 
\begin{eqnarray}
\label{eqn:2p-mgen2}
M_{k_1, k_2}\left(G_1, G_2, {\bf y}, t; s_1, s_2\right) 
= \sum_{i,j\ge 0} 
p_{k_1, k_2, i, j}(G_1, G_2, \textbf{y}, t)e^{i\,s_1 + j\,s_2}.
\end{eqnarray}

\begin{proposition}
The moment generating functions 
$M_{k_1, k_2}\left(G_1, G_2, {\bf y}, t; s_1, s_2\right)$ 
solve the following recursive system of non-linear partial differential equations:
\be
\label{eqn:2p-pdeM}
\frac{\partial}{\partial t} 	M_{k_1, k_2}
= D \Delta_{{\bf y}} M_{k_1, k_2} - \lambda M_{k_1, k_2} + \lambda \, h(M_{k_1-1, k_2-1}), 
\quad k_1, k_2 \geq 1,
\ee
with the initial conditions 
\begin{eqnarray}
\label{eqn:2p-initialM0}
&&M_{k_1, k_2}\left(G_1, G_2, {\bf y}, 0; s_1, s_2\right) 
\equiv 1, \quad k_1, k_2 \geq 1,\\
\label{eqn:2p-initialM1}
&&M_{0,0}\left(G_1, G_2, {\bf y}, t; s_1, s_2\right) 
= P_1e^{s_1}+P_2e^{s_2} + 1- P_1-P_2,\\
\label{eqn:2p-initialM2}
&&M_{0,k}\left(G_1, G_2, {\bf y}, t; s_1, s_2\right) 
= \big(M_{k} (G_2,  {\bf y}, t; s_2) - e^{-\lambda t}\big)+ (e^{-\lambda t} - P_1) +P_1e^{s_1}, 
\end{eqnarray}
where 
$P_i:=e^{-\lambda t} \int_{G_i} p({\bf x},{\bf y},t)d{\bf x}$, $i=1,2$.
Here, as before, $h(s)$ is defined by \eqref{branching_pgf} 
and $\Delta_{{\bf y}} = \sum_i \partial^2/\partial y_i^2.$
\label{prop2}
\end{proposition}
{\bf Proof} is given in Appendix~\ref{p_prop2}.

\subsubsection{Moments}
Consider the expected value $\bar A_{k_1, k_2}(G_1, G_2, \textbf{y}, t)$ of the 
product of the number of generation-$k_1$ particles in region $G_1$ and number of 
generation-$k_2$ particles in region $G_2$ at instant $t$, produced by a single 
immigrant injected at point $\textbf{y}$ at time $t=0$.  
It is given by the following partial derivative
\begin{eqnarray}
\label{eqn:2p-ex1}
\bar A_{k_1, k_2}(G_1, G_2, \textbf{y}, t) 
:= \left.\frac{\partial^2 M_{k_1, k_2} \left(G_1, G_2, \textbf{y}, t ; s_1, s_2 \right)}
{\partial s_1 \partial s_2} \right|_{s_1=s_2=0}.
\end{eqnarray}

We notice that the expectations $\bar A_{k_1}(G_1,\textbf{y}, t)$ and 
$\bar A_{k_2}(G_2, \textbf{y}, t)$ of \eqref{eqn:ex1} can be represented as
\begin{eqnarray}
\label{eqn:partial1}
	\bar A_{k_1}(G_1,\textbf{y}, t)  &:=&  \left.\frac{\partial M_{k_1, k_2}\left(G_1, G_2, \textbf{y}, t ; s_1, s_2 \right)}{\partial s_1}\right|_{s_1=s_2=0}
\end{eqnarray}
and
\begin{eqnarray}
\label{eqn:partial2}
	\bar A_{k_2}(G_2, \textbf{y}, t) &:=& \left.\frac{\partial M_{k_1, k_2}\left(G_1, G_2, \textbf{y}, t ; s_1, s_2 \right)}{\partial s_2}\right|_{s_1=s_2=0}.
\end{eqnarray}
Consider also the expectation density $A_{k_1, k_2}(\mathbf{x_1}, \mathbf{x_2}, \textbf{y}, t)$ 
that satisfies, for any nonoverlapping $G_1, G_2 \subset \mathbb R^n$ 
\begin{eqnarray}
\label{eqn:2p-ex2}
\bar A_{k_1, k_2}(G_1, G_2,  \textbf{y}, t) 
= \int_{G_2} \int_{G_1} A_{k_1, k_2}(\mathbf{x_1}, \mathbf{x_2}, \textbf{y}, t) d \mathbf{x_1} d \mathbf{x_2}.
\end{eqnarray}

\begin{corollary} 
The moment densities $A_{k_1,k_2} \equiv A_{k_1, k_2}({\bf x_1}, {\bf x_2}, {\bf y}, t)$ 
solve the following recursive system of partial differential equations:
\begin{eqnarray}
\label{eqn:2p-pdeA}
\lefteqn{\frac{\partial A_{k_1, k_2}}{\partial t} =
D \, \Delta_{{\bf y}} A_{k_1, k_2} - \lambda \, A_{k_1, k_2} + \lambda \, B \, A_{k_1-1, k_2-1}}\nonumber\\
&+& \lambda \, h^{''}(1)\,A_{k_1-1}({\bf x_1}) \, A_{k_2-1}({\bf x_2}), \quad k_1,k_2 \geq 1,
\end{eqnarray}
with the initial conditions 
\begin{eqnarray}
\label{eqn:2p-initialA1}
A_{k_1, k_2}({\bf x_1}, {\bf x_2}, {\bf y}, 0) &\equiv& 0,\quad k_1, k_2 \geq 1,\\
\label{eqn:2p-initialA2}
A_{0,k}({\bf x_1}, {\bf x_2}, {\bf y}, t) &\equiv& 0,  \quad 	k\geq 0, \: t \geq 0,
\end{eqnarray}
and $A_k({\bf x})\equiv A_k({\bf x},{\bf y},t)$ given by \eqref{eqn:solA}.
\label{col2}
\end{corollary}
{\bf Proof} is given in Appendix~\ref{p_col2}.  

\subsection{Multiple immigrants}
\label{multiple}

Here we expand the results of the Sect.~\ref{single_one} to the case of multiple 
immigrants that appear at the origin according to a homogeneous Poisson process 
with intensity $\mu$.
The expectation $\cA_k$ of the number of particles of generation 
$k$ is given by
\begin{eqnarray}
\cA_k({\bf x},t) &=& \int_0^t A_k({\bf x},{\bf 0},s)\,\mu\,ds\nonumber\\
&=&\frac{\mu\,(\lambda\,B)^k}{k!\left(4\,\pi\,D\right)^{n/2}}
\int_0^t s^{k-n/2}\,
\exp\left\{-\lambda\,s-\frac{|{\bf x}|^2}{4\,D\,s}\right\}ds.
\end{eqnarray}
The steady-state spatial distribution corresponds to the limit
$t\to \infty$:
\be
\cA_k({\bf x}):=\cA_k({\bf x},\infty)=
\frac{2\,\mu\,(\lambda\,B)^k}{k!\left(4\,\pi\,D\right)^{n/2}}
\left(\frac{|{\bf x}|^2}{4\,D\,\lambda}\right)^
{\nu/2}K_{\nu}\left(|{\bf x}|\sqrt
{\frac{\lambda}{D}}\right).
\label{A}
\ee
Here $\nu=k-n/2+1$ and $K_{\nu}$ is the modified
Bessel function of the second kind (see Appendix~\ref{Bessel}).
Introducing the normalized distance from the origin
$z:=|\vx|\sqrt{\lambda/D}$ we obtain
\be
\cA_k(z)=\frac{\mu}
{\lambda\,k!}\left(\frac{B}{2}\right)^k
\left(\frac{2\,\pi\,D}{\lambda}\right)^{-n/2}\,
z^{\nu}\,K_{\nu}(z).
\label{Az}
\ee
For odd $n$, there are explicit expressions for $K_{\nu}(z)$
(Appendix~\ref{Bessel}, Eqs.~\eqref{K1},\eqref{K2}).
In particular, we have 
\be
\cA_0(z)=\frac{\mu}{\sqrt{4\,D\,\lambda}}
e^{-z},{\rm ~for~}n=1,\quad
\cA_0(z)=\sqrt{\frac{\lambda}{D^3}}\,\frac{\mu}{4\,\pi\,z}
e^{-z}, {\rm ~for~}n=3.
\ee 
From \eqref{Az} and the asymptotic behavior of $K_{\nu}(z)$ as $z\to 0$ 
(Appendix~\ref{Bessel}, Eq.~\eqref{lim0})  
it follows that 
\be
\displaystyle\lim_{z\to 0}\,\cA_k(z)=\left\{
\begin{array}{cc}
\infty,& {\rm for~} \nu \le 0, {~i.e.,~} k\le n/2-1\\
const<\infty,& {\rm for~} \nu >0, {~i.e.,~} k > n/2-1.
\end{array}
\right.
\label{as}
\ee
Thus, in a model with spatial dimension $n\ge 2$, the elements of 
several lowest generations $(k\le n/2-1)$ have an infinite concentration 
at the origin.

\subsection{Alternative model representation}
In this section we derive a system of equations for the steady-state expectations
$\cA_k({\bf x})$ using the radial symmetry of the problem. 
By integrating the equation \eqref{eqn:pdeA} from $t = 0$ to $\infty$, we obtain
\[D \Delta_{\mathbf{x}} \mathcal A_k({\bf x}) - \lambda \mathcal A_k({\bf x}) + \lambda B \mathcal A_{k-1}({\bf x})=0\]
since $A_k(\mathbf x, \mathbf y, \infty)=0$.  
We now rewrite this equation in terms of the normalized distance from the origin, 
$z: = |\mathbf x| \sqrt{\lambda/D}$, using the fact that 
$\cA_k({\bf x})\equiv\cA_k(z)$ as soon as $|{\bf x}|=|z|$:
\begin{equation}
\label{eqn:mathcalAdeq}
\mathcal A_k^{''}(z) + \frac{n-1}{z}	\; \mathcal A_k^{'}(z)- 	\mathcal A_k(z) + B \mathcal A_{k-1}(z) = 0.   
\end{equation}

We notice, furthermore, that one can rewrite the expectation densities \eqref{eqn:solA}
as a function of $z$, which results in $A_k(z)\equiv A_k({\bf x},{\bf 0},t)$.
It is then readily seen that
\be
A_k^{'}(z) =  - \frac{B}{2k} \; z \; A_{k-1}(z).
\label{simple}
\ee
The same recursive system holds for $\mathcal A_k(z)$, which is shown by integrating
the last equation with respect to time.

\section{Particle rank distribution}
\label{GR}

We analyze here the particle rank distribution; recall that the rank is defined 
as $r=r_{\rm max}-k$, where $k$ is the particle's generation.
A self-similar branching mechanism that governs our model suggests an exponential distribution 
of particle ranks.
Indeed, the spatially averaged steady-state rank distribution is a pure exponential law with 
index $B$:
\begin{eqnarray}
A_k:&=&\int_{\bR^n}\int_0^{\infty} 
A_k({\bf x,0},t) \mu\,dt\,d{\bf x}\nonumber\\
&=&\frac{\mu\,B^k}{k!}\int_0^{\infty}(\lambda\,t)^k
\,e^{-\lambda\,t}\,dt=\frac{\mu}{\lambda}\,B^k\propto B^{-r}.
\label{pureexp}
\end{eqnarray}

\begin{remark}
Our use of the term ``self-similar'' with respect to the exponential distribution,
often seen in physical literature, requires some explanations.
As we mentioned earlier, the particle rank serves as a logarithmic measure of
its size.
Thus, the exponential distribution of ranks corresponds to the power law 
distribution of sizes; hence the term ``self-similarity''.  
\end{remark}
To analyze rank- and space-dependent deviations from the pure exponential distribution, we will 
consider the ratio $\gamma_k({\bf x})$ between the number of particles of two 
consecutive generations:
\be
\gamma_k(\vx):=\frac{\cA_k(\vx)}{\cA_{k+1}(\vx)}.
\label{gamma}
\ee
For the purely exponential rank distribution, $A_k({\bf x}) = c\,B^k$, the 
value of $\gamma_k({\bf x})=1/B$ is independent of $k$ and ${\bf x}$; 
while deviations from the pure exponential distribution will cause $\gamma_k$ to
vary as a function of $k$ and/or ${\bf x}$. 
Plugging (\ref{Az}) into (\ref{gamma}) we find
\be
\gamma_k(\vx)=\frac{2\,(k+1)}{B\,z}\,
\frac{K_{\nu}(z)}{K_{\nu+1}(z)},
\label{gamma1}
\ee
where, as before, $z:=|\vx|\,\sqrt{\lambda/D}$ and $\nu=k-n/2+1$.

\begin{proposition}
The asymptotic behavior of the function $\gamma_k(z)$ is given by
\begin{eqnarray}
\lim\limits_{z\to 0}\gamma_k(z)&=&\left\{
\begin{array}{cc}
\infty,&\nu\le 0,\\
\displaystyle\frac{1}{B}\left(1+\frac{n}{2\,\nu}\right),&
\nu>0,
\end{array}\right.
\label{gammaz0}\\
\gamma_k(z)&\sim&
\frac{2(k+1)}{B\,z},\quad {z\to\infty},\quad {\rm fixed~}k,
\label{gammazinf}\\
\gamma_k(z)&\sim&
\frac{1}{B}\left(1+\frac{n}{2\,\nu}\right),\quad {k\to\infty},
\quad {\rm fixed~}z.
\label{gammakinf}
\end{eqnarray}
\label{gammalim}
\end{proposition}
{\bf Proof} and explicit rates of divergence in \eqref{gammaz0}
are given in Appendix~\ref{p_gammalim}. 

Proposition \ref{gammalim} describes the spatio-temporal deviations of 
the particle rank distribution from the pure exponential law \eqref{pureexp}.
We interpret below each of the equations \eqref{gammaz0}-\eqref{gammakinf}
in some detail.  
Eq.~\eqref{gammakinf} implies that at any spatial point, the 
distribution asymptotically approaches the exponential form as generation
$k$ increases (rank $r$ decreases).
In other words, the distribution of small ranks (large generation 
numbers) is close to the exponential with index $-B$; thus the
deviations can only be observed at the largest ranks (small generation
numbers).
Analysis of the large-rank distribution is done using 
Eqs.~\eqref{gammaz0} and~\eqref{gammazinf}. 
Near the origin, where the immigrants enter the system,
Eq.~\eqref{gammaz0} implies that $\gamma_k(z) > \gamma_{k+1}(z)>1/B$ for $\nu >0$.
Hence, one observes the {\it upward deviations} from the pure exponential distribution:
for the same number of rank $r$ particles, the number of
rank $r+1$ particles is larger than predicted by the exponential law.
The same behavior is in fact observed for $\nu\le 0$ 
(see Appendix~\ref{p_gammalim}, Eq.~\eqref{gammalims}).
In addition, for $\nu\le 0$ the ratios $\gamma_k(z)$ do not merely
deviate from $1/B$, but diverge to infinity at the origin.
Away from the origin, according to Eq.~\eqref{gammazinf}, we have 
$\gamma_k(z)<\gamma_{k+1}(z)<1/B$, which implies {\it downward deviations}
from the pure exponent: for the same number of rank $r$ particles,
the number of rank $r+1$ particles is smaller than predicted by the exponential law.

Figure~\ref{fig_Ak} illustrates the above findings;
it shows the distribution of particles for the largest ranks
at different distances from the origin. 
One can clearly see the transition from downward to upward
deviation of the rank distributions from the pure exponential form as 
we approach the origin. 
Notably, the magnitude of the upward deviation close to the origin
(the upper line in all panels) strongly increases with the model 
dimension $n$.

\section{Numerical analysis}
\label{num}
Our analytical results and asymptotics are closely reproduced in numerical experiments 
with finite number of generations, limited spatial extent, and spatial averaging 
(unavoidable when working with observations). 
Here, to mimic the ensemble averaging, the numerical results have been averaged over 
4000 independent realizations of a 3D model with parameters $\mu=\lambda=1$, $D=1$, 
and $B=2$.

First, we check the exponential rank distribution of \eqref{pureexp}. 
Figure~\ref{fig_GR} shows the observed spatially averaged particle rank distribution.
The exponential form \eqref{pureexp} is indeed well reproduced.

Next, we see how the spatial averaging affects the rank distribution.
Figure~\ref{fig_GR_3D} shows the rank distribution at $t=30$ at various
distances to the origin.
The spatial averaging has been done within spherical shells (space between
two concentric spheres) of a constant volume $V=5$.
Thus, here we see an observable counterpart of the theoretical
distributions shown in Fig.~\ref{fig_Ak}b.
Although the spatial averaging somewhat tapers off the upward
bend at the largest ranks close to the origin,
the predicted transition from the downward to upward bend is 
clearly seen.

\begin{figure}
\centering\includegraphics[width=.8\textwidth]{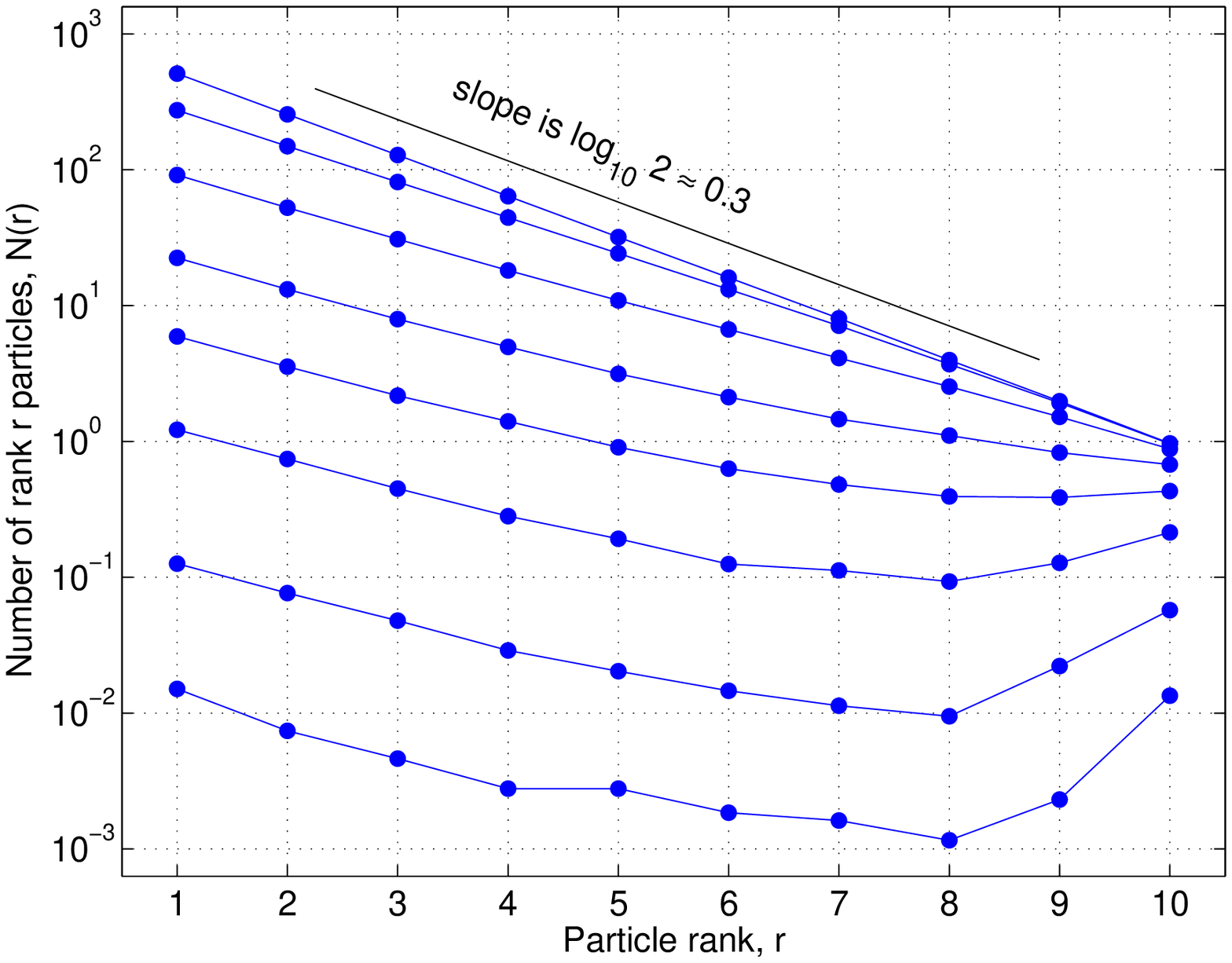}
\caption{Particle rank distribution at $t=30$ in a 3D model. 
The distribution is spatially averaged over spheres of volume $V$ 
centered at the origin with (from top to bottom):
$V=\infty, 500, 100, 200, 5, 1, 0.01$. 
Model parameters: $\mu=\lambda=1$, $D=1$, 
$B=2$, $r_{\rm max}=10$. 
The upward deviations from the exponential distribution (a straight line)
are fading away with the extent of the spatial averaging.
}
\label{fig_GR_dist}
\end{figure}

Figure~\ref{fig_GR_dist} illustrates in more detail how the
spatial averaging affects the upward bend in a 3D model.
It shows the particle rank distributions
at $t=30$ spatially averaged over spheres of different volumes
centered at the origin.
The upward bend is prominent for the spheres with volumes $V\le 5$;
and it gradually disappears within larger spheres in favor of an 
exponential distribution observed after a complete spatial averaging.
Notably, the pure exponential distribution can be only achieved
by averaging over {\it all} events in the model ($V=\infty$).
 
\section{Discussion}
\label{discussion}
This work is motivated by the problem of prediction of extreme events in 
complex systems. 
Our point of departure is the four types of premonitory patterns \cite{KB02},
previously found in models and observations.
We propose here a simple mechanism and a single control parameter for 
all these patterns.

Quantitative analysis is performed here for a classical model of spatially 
distributed population of particles of different sizes governed by direct 
cascade of branching and external driving (see Sect.~\ref{model}). 
In the probability theory this model is known as the age-dependent multitype 
branching diffusion process with immigration \cite{AN04}. 
We consider here a new scope of problems for this model. 
We assume that observations (detection of particles) are only possible on a 
subspace of the system space while the source of external driving (origin) 
remains unobservable, as is the case in many real-world systems. 
The natural question under this approach is the dependence of the process 
statistics on the distance to the source. 
A complete analytical solution to this problem, in terms of the moments with 
respect to the particle density, is given by Proposition~\ref{prop1}. 
In addition, the correlation structure of the particle field can be found 
using Proposition~\ref{prop2}.

It is natural to consider rank as a logarithmic measure of the particle size. 
The exponential rank distribution 
derived in Eq.~\eqref{pureexp} corresponds to a self-similar, power-law distribution of 
particle sizes, characteristic for many complex systems. 
The self-similarity in our model, as well as in the real-world systems, is only observed 
after global spatial averaging in a steady-state. 
Proposition ~\ref{gammalim} and Fig.~\ref{fig_Ak} describe space-dependent deviations from the 
self-similarity. 
Recall that an extreme event in our system is defined as an observation of a 
particle of sufficiently large size. 
As the source approaches the observation subspace, the probability of an extreme 
event increases. 
Our results are thus directly connected to prediction: When the location of the source 
changes in time approaching the subspace of observation (or {\it vice versa}), 
the increase of event intensity and the downward bend in the event size distribution 
becomes premonitory to an extreme event. 
The numerical experiments confirm the validity of our analytical results and 
asymptotics in a finite model.

Our model exhibits very rich and intriguing premonitory behavior. 
Figure ~\ref{fig_example} shows several 2D snapshots of a 3D model at different 
distances from the source. 
One can see that, as the source approaches, the following changes in the background 
activity emerge:
a) The intensity (total number of particles) increases;
b) Particles of larger size become relatively more numerous;
c) Particle clustering becomes more prominent;
d) The correlation radius increases.
All these premonitory changes have been independently observed in natural and 
socioeconomic systems. 
Here they are all determined by a single control parameter -- distance between
the source and the observation space.

The abovementioned premonitory patterns closely resemble universal properties 
of models of statistical physics in a vicinity of 
second order phase transition \cite{Stanley71,Ma00,Kadanoff00}, 
percolation models near the percolation threshold \cite{SA94,Grimmett}, 
or random graphs prior to the emergence of a giant cluster \cite{Bollobas,Durrett,NBW06}. 
In these models, the approach of an extreme event, usually referred to as critical point, 
and the emergence of premonitory patterns, called critical phenomena, correspond to 
an instant when a control parameter crosses its critical value. 
In statistical physics a typical control parameter is temperature or magnetization; 
in percolation it is the site or bond occupation density; in a random graph --- 
the probability for two vertices to be connected. 
The theory of critical phenomena \cite{Ma00} quantifies system's behavior at 
the critical value of the corresponding control parameter. 
The remarkable power of this theory is connected to the fact that very different 
systems demonstrate similar behavior near to criticality. 
More precisely, when the control parameter is close to its critical value, the system 
sticks to one of just a few types of possible limit behaviors, each being described by 
an appropriate scale-invariant statistical field theory. 
In particular, each limit behavior corresponds to the asymptotic power-law size 
distribution of system observables with a characteristic value of critical exponent.

We focus here on a problem inverse to that considered by the critical phenomena theory: 
Estimating the deviation of a control parameter from the critical value using 
the observed system behavior. 
The motivation for this is coming from environmental, geophysical, and other applied 
fields where one faces a problem of assessing the likelihood of occurrence of an 
extreme event associated with a critical point. 
We formulate and solve such a prediction problem for a spatially embedded cascade process, 
which enjoys both the mean-field self-similarity and realistic premonitory time- and 
space-dependent deviations from the latter.
The methods developed in this paper may provide a framework for studying predictability 
of extreme events in complex systems of arbitrary nature.
 
\acknowledgments
This research was partly supported by NSF grants ATM-0620838
and EAR-0934871 (to IZ) and DMS-0801050 (to AG).

\appendix

\section{Proof of Proposition~\ref{prop1}}
\label{p_prop1}

We will need the following calculus lemma that is readily proven
by using the definition of derivative:
\begin{lemma}
Let $f(z)$, $g(z)$, $z\in\mathbb{R}$ be continuous functions such that 
the definite integral
$G(t)=\int_0^t f(z)\,g(t-z)dz$ exists. 
We also assume that $g(z)$ is differentiable.
Then,
\[\frac{d}{dt}G(t)=\int_0^t f(z)\,g'(t-z)dz+f(t)\,g(0).\]
\label{lemma}
\end{lemma} 

There are two possible scenarios for the model development up to time $t$.  
In the first one, the initial immigrant will not split; the probability for 
this is $P = e^{-\lambda t}$.  
In the second one, the initial immigrant will split at instant $0 \leq u \leq t$; 
the probability of the first split within the time interval $[u, u+du]$ is 
$\lambda e^{-\lambda t}du + o(du)$ as $du \rightarrow 0$.  
The spatial position of the split is given by the diffusion density 
$p({\bf x},{\bf y},u)$.  
If the immigrant splits, the composition property of generating functions 
 gives $M_k = h[M_{k-1}]$.  
Integrating over all possible split instants and locations, we obtain
\begin{eqnarray}
\label{eqn:aggregate}
M_k(G,{\bf y},t;s)
=e^{-\lambda t}+ \int_{\mathbb R^n} d\textbf{y}' \int_0^t du \, 
\lambda e^{-\lambda u} p(\textbf{y}', \textbf{y}, u) \, 
h[M_{k-1}(G,\textbf{y}', t-u; s)].
\end{eqnarray}
Here the first and the second terms correspond to the first and second scenarios, respectively.  
Using the new integration variable $z = t-u$, we write
\begin{eqnarray*}
M_k(G,{\bf y},t;s) 
&=& e^{-\lambda t}+ e^{-\lambda t}\int_{\mathbb R^n} d\textbf{y}' 
\int_0^t du \, \lambda e^{\lambda(t-u)} p(\textbf{y}', \textbf{y}, u) \,
h[M_{k-1}(G,\textbf{y}', t-u; s)] \\
&=& e^{-\lambda t}\left(1+ \int_{\mathbb R^n} d\textbf{y}' \int_0^t dz \, 
\lambda e^{\lambda z} p(\textbf{y}', \textbf{y}, t-z) \; h[M_{k-1}(G, \textbf{y}', z; s)]\right).
\end{eqnarray*}

Now we take the derivative with respect to $t$ of both sides and apply Lemma A.1 using the fact that 
$p(\textbf{y}', \textbf{y}, 0)=\delta(\textbf{y}'-\textbf{y})$ and $(\partial/\partial t - D \Delta_{\textbf{y}})p = 0:$
\begin{multline*}
	\frac{\partial}{\partial t} 	M_k(G,  \textbf{y}, t ;s) = -\lambda M_k (G,  \textbf{y}, t ;s)\\
	+ e^{-\lambda t}\left[\int_{\mathbb R^n} d\textbf{y}' \int_0^t dz \, \lambda e^{\lambda z} \; 
	h[M_{k-1}(G, \textbf{y}', z; s)]D \Delta_{\textbf{y}} p(\textbf{y}', \textbf{y}, t-z) +   
\lambda e^{\lambda t} \; h[M_{k-1}(G,  \textbf{y}, t ;s)]  \right].
\end{multline*}
Taking the operator $\Delta_{\textbf{y}}$ out of the integration signs, we find
\[	\frac{\partial}{\partial t} 	M_k(G,  \textbf{y}, t ;s) 
= D\Delta_{\textbf{y}} M_k - \lambda M_k + \lambda \; h[M_{k-1}].\]
It is left to establish the initial conditions.  
Since we start the model with a particle of generation $k=0$ and the distribution of splitting 
is continuous, at $t=0$ there are no other particles with probability 1.  
Hence, $M_k(G,  \textbf{y}, 0 ;s)=1$ for all $k \geq 1$.  
For generation $k=0$, we can only have one or no particles at time $t >0$.  
The probability to have one particle is given by the product of probabilities that there was 
no split up to time $t$ and that the particle happens to be within region $G$ at time $t$: 
$P = e^{-\lambda t} \int_G p(\textbf{x}, \textbf{0}, t) d\textbf{x}$.  
The probability to have no particles is then $(1-P)$.  
This implies (\ref{eqn:initialM}).   \hfill
$\Box$

\section{Moments in one-point system}
\label{p_col1}
For any natural number $j$, consider the $j$-th moment $\bar A^{(j)}_k(G, {\bf y}, t)$ of 
the number of generation-$k$ particles at instant $t$ within the region $G$, produced by a 
single immigrant injected at point ${\bf y}$ at time $t=0$.  
It is given by the following partial derivative (see {\it e.g.}, \cite{AN04}, Chapter 1):
\begin{eqnarray}
\label{eqn:ex1a}
\bar A^{(j)}_k(G,{\bf y}, t) := \frac{\partial^j M_k(G,{\bf y}, t ; s)}{\partial s^j} \Big|_{s=0}.
\end{eqnarray}

\begin{corollary}
The moments $\bar A_k^{(j)}(G,{\bf y}, t)$  
solve the following recursive system of partial differential equations:
\begin{equation}
\label{eqn:pdeA-j}
	\frac{\partial}{\partial t} \bar A_k^{(j)}(G, \mathbf y, t) =
D \Delta_{\mathbf y} \bar A_k^{(j)} - \lambda \bar A_k^{(j)} + \lambda 
\left[\sum \frac{j!}{m_1! m_2! \dots m_j!}h^{(m)}(1) \prod_{i=1}^j \left(\frac{\bar A_{k-1}^{(i)}}{i!}\right)^{m_i}\;\right], 
\end{equation} 
where $m=m_1+\dots +m_j$ and the sum is over all partitions of $j$, i.e., values of $m_1, \ldots, m_j$ such that
$m_1+2m_2+ \dots +jm_j=j$,
with the initial conditions 
\begin{eqnarray}
\label{eqn:initialA-j1}
\bar A_k^{(j)}(G, {\bf y}, 0) &\equiv& 0, k \geq 1,\\
\label{eqn:initialA-j2}
\bar A_0^{(j)}(G, {\bf y}, 0) &=& \int_G \delta(\mathbf{y}-\mathbf{x}) d{\bf x}, \\	
\label{eqn:initialA-j3}
\bar A_0^{(j)}(G, {\bf y}, t) &=& e^{-\lambda t} \int_G p(\mathbf{x}, \mathbf{y}, t) d{\bf x}, \; t> 0,
\end{eqnarray}
and
\[h^{(i)}(1):= \left.\frac{d^i}{ds^i}h(s)\right|_{s=1}=\sum_{n=i}^\infty \frac{n!}{(n-i)!} \,p_n.\] 
\label{col1}
\end{corollary}
{\bf Proof:}
The validity of \eqref{eqn:pdeA-j} follows from Proposition~\ref{prop1}.  
Namely, applying the operator $\partial^j/\partial s^j( \cdot)|_{s=0}$ to both 
sides of \eqref{eqn:pdeM}, changing the order of differentiation, and using 
Fa\`a di Bruno's formula for the $j$-th derivative of a composition 
function, one finds, for each $k \geq 1$, 
\begin{eqnarray*}
\frac{\partial^j}{\partial s^j} \left.\left[\frac{\partial M_k (G,\textbf{y}, t;s)}{\partial t} \right]\right|_{s=0}
&=& 
\frac{\partial^j}{\partial s^j} \left[D \Delta_{\bf{y}} M_k - \lambda M_k + \lambda \, h(M_{k-1}) \right]|_{s=0},\\
\frac{\partial}{\partial t} \left. \left[\frac{\partial^j M_k (G,\textbf{y}, t;s)}{\partial s^j}  \right|_{s=0} \right]
&=& 
\left. \left[D \Delta_{\bf{y}} \frac{\partial^j M_k}{\partial s^j}  - \lambda \frac{\partial^j M_k}{\partial s^j}  + 
\lambda \, \frac{\partial^j}{\partial s^j} h(M_{k-1}) \right]\right|_{s=0},
\end{eqnarray*}
\begin{eqnarray*}
\lefteqn{ \frac{\partial}{\partial t}\bar A^{(j)}_k \left(G,{\bf y}, t\right) 
= \left[D \Delta_{{\bf y}} \frac{\partial^j M_k}{\partial s^j}  - 
\lambda \frac{\partial^j M_k}{\partial s^j}\right.} \\
&+& \left. \left. 
\lambda \, \left(\sum \frac{j!}{m_1! m_2! \dots m_j!} \, h^{(m)}(M_{k-1}) 
\prod_{i=1}^j \left(\frac{M_{k-1}^{(i)}}{i!}\right)^{m_i} \right)\right]\right|_{s=0} \\ 
&=& D \Delta_{\bf y} \bar A_k^{(j)} - \lambda \bar A_k^{(j)} + \lambda 
\left[\sum \frac{j!}{m_1! m_2! \dots m_j!}h^{(m)}(1) \prod_{i=1}^j 
\left(\frac{\bar A_{k-1}^{(i)}}{i!}\right)^{m_i}\;\right], 
\end{eqnarray*} 
where $m=m_1+\dots +m_j$ and the sum is over all partitions of $j$, i.e., values of $m_1, \ldots, m_j$ such that
$m_1+2m_2+ \dots +jm_j=j$.  
The initial conditions are established by applying the operator 
$\partial^j/\partial s^j( \cdot)|_{s=0}$ to 
both sides of \eqref{eqn:initialM} and using the definition of $\bar A^{(j)}_{k}(G, \mathbf{y}, t)$ 
in \eqref{eqn:ex1}.  

\section{Proof of Corollary~\ref{col11}}
\label{p_col11}
For $j=1$, the equation in Corollary~\ref{col1} simplifies to 
\[\frac{\partial}{\partial t}\bar A_k (G,\textbf{y}, t) 
= D \Delta_{\textbf{y}} \bar A_k - \lambda \bar A_k + \lambda B \bar A_{k-1}.\]
Using the definition of $A_k(\textbf{x}, \textbf{y}, t) $ given in (\ref{eqn:ex2}), 
one obtains for each $k \geq 1$, 
\[
\frac{\partial}{\partial t}  A_k (\textbf{x},\textbf{y}, t) 
= D \Delta_{\textbf{y}}  A_k - \lambda A_k + \lambda B  A_{k-1}.  
\]
It is left to use the translation property 
$A_k(\textbf{x}, \textbf{y}, t)=A_k(\textbf{x}-\textbf{y}, \textbf{0}, t)$ 
to change $\Delta_{\textbf{y}}$ to $\Delta_{\textbf{x}}$.  

The validity of general solution \eqref{eqn:solA} is proven by induction using the fact that 
\[\left[\frac{\partial}{\partial t} - D \Delta_{\textbf{x}} + \lambda\right] A_0 = 0.\]
The last equality in \eqref{eqn:solA} follows from 
\eqref{eqn:initialA-j3} and \eqref{sol}.   
\qed

\section{Proof of Proposition~\ref{prop2}}
\label{p_prop2}

The proof of Proposition~\ref{prop2} follows the line of the proof of Proposition~\ref{prop1}.  
There are two possible scenarios for the model development up to time $t$.  
In the first one, the initial immigrant will not split; the probability for this is $P = e^{-\lambda t}$.  
In the second one, the initial immigrant will split at instant $0 \leq u \leq t$; 
the probability of the first split within the time interval $[u, u+du]$ is 
$\lambda e^{-\lambda t}du + o(du)$ as $du \rightarrow 0$.  
The spatial position of the split is given by the diffusion density $p(\textbf{x},  \textbf{y}, u)$.  
If the immigrant splits, the composition property of generating functions gives 
$M_{k_1, k_2} = h[M_{k_1-1, k_2-1}]$.  
Integrating over all possible split instants and locations, we obtain
\begin{multline*}
M_{k_1, k_2}\left(G_1, G_2, \textbf{y}, t; s_1, s_2\right) \\
= e^{-\lambda t}+ \int_{\mathbb R^n} d\textbf{y}' \int_0^t du \, 
\lambda e^{-\lambda u} p(\textbf{y}', \textbf{y}, u) \, 
h\left[M_{k_1-1, k_2-1}\left(G_1, G_2, \textbf{y}', t-u; s_1, s_2\right)\right].
\end{multline*}
Here the first and the second terms correspond to the first and second scenarios, respectively.  
Using the new integration variable $z = t-u$, we write
\begin{multline*}
M_{k_1, k_2}\left(G_1, G_2, \textbf{y}, t; s_1, s_2\right)  \\
= e^{-\lambda t}+ e^{-\lambda t}\int_{\mathbb R^n} d\textbf{y}' 
\int_0^t du \, \lambda e^{\lambda(t-u)} p(\textbf{y}', \textbf{y}, u) \,
h\left[M_{k_1-1, k_2-1}\left(G_1, G_2, \textbf{y}', t-u; s_1, s_2\right)\right] \\
= e^{-\lambda t}\left(1+ \int_{\mathbb R^n} d\textbf{y}' 
\int_0^t dz \, \lambda e^{\lambda z} p(\textbf{y}', \textbf{y}, t-z) \, 
h\left[M_{k_1-1, k_2-1}\left(G_1, G_2, \textbf{y}', z; s_1, s_2\right)\right]\right).
\end{multline*}
Now we take the derivative with respect to $t$ of both sides and apply Lemma~\ref{lemma}
using the fact that 
$p(\textbf{y}', \textbf{y}, 0)=\delta(\textbf{y}'-\textbf{y})$ and 
$(\partial/\partial t - D \Delta_{\textbf{y}})p = 0$:

\begin{multline*}
\frac{\partial}{\partial t} 	
M_{k_1, k_2}\left(G_1, G_2, \textbf{y}, t; s_1, s_2\right)  
= -\lambda M_{k_1, k_2}\left(G_1, G_2, \textbf{y}, t; s_1, s_2\right) \\
+ e^{-\lambda t}\left[\int_{\mathbb R^n} d\textbf{y}' \int_0^t dz \, \lambda e^{\lambda z} \, 
h\left[M_{k_1-1, k_2-1}\left(G_1, G_2, \textbf{y}', z; s_1, s_2\right)\right]
D \Delta_{\textbf{y}} p\left(\textbf{y}', \textbf{y}, t-z\right) \right. \\
+  \left.\lambda e^{\lambda t} \; h\left[M_{k_1-1, k_2-1}
\left(G_1, G_2,  \textbf{y}, t ;s_1, s_2\right)\right]\vphantom{\int_0^0}\right]. 
\end{multline*}
Taking the operator $\Delta_{\textbf{y}}$ out of the integration signs, we find
\[	\frac{\partial}{\partial t} 	M_{k_1, k_2}\big(G_1, G_2, \textbf{y}, t; s_1, s_2\big) 
= D\Delta_{\textbf{y}} M_{k_1, k_2} - \lambda M_{k_1, k_2} + \lambda \; h[M_{k_1-1, k_2-1}].\]

It is left to establish the initial conditions.  
Since we start the model with a particle of generation $k=0$ and the distribution of splitting 
is continuous, at $t=0$ there are no other particles with probability 1.  
Hence, $M_{k_1, k_2}\big(G_1, G_2, \textbf{y}, 0; s_1, s_2\big) =1$ 
for all $k_1, k_2 \geq 1$.  
For generation $k_1=k_2=0$, we have three possibilities: the initial immigrant has not split 
and is in $G_1$ ($i=1, j=0$),
the initial immigrant has not split and is in $G_2$ ($i=0, j=1$), 
and neither ($i=0, j=0$), with corresponding probabilities of 
$P_1$, $P_2$, and $1-P_1-P_2$, respectively.  
This implies \eqref{eqn:2p-initialM1}.  

For generation $k_1=0$ and $k_2 = k \geq 1$, we again have three possibilities: 
the initial immigrant has not split and is in $G_1$ ($i=1, j=0$), 
the initial immigrant has not split and is not in $G_1$ ($i=0, j=0$), 
and the initial immigrant has split ($i=0, j\geq 0$), 
with corresponding probabilities of $P_1$, $e^{-\lambda t} -P_1$, and $1-e^{-\lambda t}$, respectively.    
In the last case, the number of the $0$-th generation particles in $G_1$ is $0$ with probability 1 while 
the information on the $k$-th generation particles in $G_2$ is given by 
\[\int_{\mathbb R^n} d\textbf{y}' \int_0^t du \, \lambda e^{-\lambda u} 
p(\textbf{y}', \textbf{y}, u) \; h[M_{ k-1}(G_2, \textbf{y}', t-u; s_2)].\]
From \eqref{eqn:aggregate}, we see that the above expression equals 
$M_k(G_2, \textbf{y}, t; s_2) - e^{-\lambda t}.$
This implies \eqref{eqn:2p-initialM2}.  
We notice that setting $s_2 = 0$ in \eqref{eqn:2p-initialM1} and 
\eqref{eqn:2p-initialM2} each yields $(1-P_1) + P_1 e^{s_1}$ as it should (cf. \eqref{eqn:initialM}).  
\qed

\section{Proof of Corollary~\ref{col2}}
\label{p_col2}

The validity of \eqref{eqn:2p-pdeA} follows from Proposition~\ref{prop2} and the 
definition of $\bar A_{k_1, k_2}(G_1, G_2, \mathbf{y}, t)$,
$A_{k_1, k_2}({\bf x_1}, {\bf x_2}, \mathbf{y}, t)$.
Formally, applying the operator $\partial^2/\partial s_1 \partial s_2( \cdot)|_{s_1=s_2=0}$ 
to both sides of \eqref{eqn:2p-pdeM} and changing the order of differentiation, 
one finds, for each $k_1, k_2 \geq 1$, 
\be
\frac{\partial^2}{\partial s_1\partial s_2} 
\left[\frac{\partial  M_{k_1, k_2}}{\partial t}\right]\Big|_{s_1=s_2=0} 
= 
\frac{\partial^2}{\partial s_1\partial s_2} 
\left[D \Delta_{\bf{y}} M_{k_1, k_2} - \lambda M_{k_1, k_2} 
+ \lambda \, h(M_{k_1-1, k_2-1}) \right]\Big|_{s_1=s_2=0},
\nonumber
\ee
\be
\frac{\partial}{\partial t} 
\left[\frac{\partial^2 M_{k_1, k_2}}{\partial s_1\partial s_2}  \Big|_{s_1=s_2=0}  \right]
= 
\left[D \Delta_{\bf{y}} \frac{\partial^2 M_{k_1, k_2}}{\partial s_1\partial s_2}  
- \lambda \frac{\partial^2 M_{k_1, k_2} }{\partial s_1\partial s_2} +  
\lambda \, h^{'}(M_{k_1-1, k_2-1})\frac{\partial^2 M_{k_1-1, k_2-1}}{\partial s_1\partial s_2} \right.
\nonumber
\ee
\be
\qquad \left.  +  \lambda h^{''}(M_{k_1-1, k_2-1}) 
\frac{\partial M_{k_1-1, k_2-1}}{\partial s_1}\,
\frac{\partial M_{k_1-1, k_2-1}}{\partial s_2}\right]\Big|_{s_1=s_2=0},
\nonumber
\ee
\be
\frac{\partial}{\partial t} 
 = D \Delta_{\bf{y}} \bar A_{k_1, k_2} - \lambda \bar A_{k_1, k_2} + \lambda B \bar A_{k_1-1, k_2-1} 
\nonumber
+ \lambda h^{''}(1) \bar A_{k_1-1}(G_1) \bar A_{k_2-1}(G_2).	
\ee
The system \eqref{eqn:2p-pdeA} readily follows now from the definition of 
$A_{k_1, k_2}({\bf x_1}, {\bf x_2}, \mathbf{y}, t)$.
The initial conditions (\ref{eqn:2p-initialA1}) - (\ref{eqn:2p-initialA2}) are established by 
applying the operator $\partial^2/\partial s_1 \partial s_2( \cdot)|_{s_1=s_2 =0}$ to 
both sides of \eqref{eqn:2p-initialM0} - \eqref{eqn:2p-initialM2} 
and using again the definition of $A_{k_1, k_2}({\bf x_1}, {\bf x_2}, \mathbf{y}, t)$.
\qed

\section{Proof of Proposition~\ref{gammalim}}
\label{p_gammalim}
The asymptotic \eqref{gammazinf} readily follows from \eqref{K3}.
To prove \eqref{gammakinf}, 
let $r_{\nu}(z):=K_{\nu}(z)/K_{\nu+1}(z)$.
From \eqref{K4} one finds that
\be
\frac{K_{\nu+1}(z)}{K_{\nu}(z)}=\frac{K_{\nu-1}(z)}{K_{\nu}(z)}+
\frac{2\,\nu}{z}
\label{rec}
\ee
and furthermore
\be
\frac{z}{2\,\nu}\frac{1}{r_{\nu}(z)}=
\frac{z}{2\,\nu}{r_{\nu-1}(z)}+1.
\label{r}
\ee
From monotonicity of $K_{\nu}(z)$ with respect to the index $\nu>0$ 
it follows that $r_{\nu}(z)<1$ for $\nu>0$.
Accordingly, the first term in the rhs of \eqref{r} goes to zero
as $k\to\infty$.
Hence,
\be
\lim\limits_{k\to\infty}\frac{z}{2\,\nu}\frac{1}{r_{\nu}(z)}=1,\quad
{\rm or}\quad r_{\nu}(z)\sim \frac{z}{2\,\nu},\quad k\to\infty.
\ee
To complete the proof of \eqref{gammakinf}, we use this asymptotic in \eqref{gamma1}.
Finally, we prove \eqref{gammaz0}.
In fact, we will derive a stronger result showing the asymptotics
of $r_{\nu}(z)$ and $\gamma_{\nu}(z)$ as $z\to 0$.
To find the asymptotics for $r_{\nu}(z)$, we use \eqref{lim0}
for all possible combinations of signs for $\nu$ and $\nu+1$.
We take into account that by definition $\nu$ can only take 
values $\{i,i+1/2\}_{i\in\mathbb{Z}}$.
\be
r_{\nu}(z)=\left\{\begin{array}{llllll}
\frac{K_{\nu}(z)}{K_{\nu+1}(z)}&\sim&
\frac{\Gamma(-\nu)}{\Gamma(-\nu-1)}
\left(\frac{2}{z}\right)^{-\nu-(-\nu-1)}&\sim&
2\,(-\nu-1)/z,&\nu\le -3/2,\\
\frac{K_{-1}(z)}{K_0(z)}&\sim&
[z\,(\ln(2/z)-\gamma)]^{-1}&\sim&
\displaystyle-(z\,\ln\,z)^{-1},&\nu=-1,\\
\frac{K_{-1/2}(z)}{K_{1/2}(z)}&&&=&1,&\nu=-1/2,\\
\frac{K_{0}(z)}{K_{1}(z)}&\sim&
z\,(\ln(2/z)-\gamma)&\sim& -z\,\ln\,z,&\nu=0,\\
\frac{K_{\nu}(z)}{K_{\nu+1}(z)}&\sim&
\frac{\Gamma(\nu)}{\Gamma(\nu+1)}
\left(\frac{2}{z}\right)^{\nu-(\nu+1)}&=&
z/(2\,\nu),&\nu>0.
\end{array}\right.
\ee
Combining this with \eqref{gamma1} we find
\be
\gamma_{\nu}(z)=\frac{2\,(k+1)}{B\,z}\,r_{\nu}(z)\sim\left\{\begin{array}{cl}
\frac{4}{B\,z^2}(\nu+n/2)\,(-\nu-1),&\nu\le-3/2,\\
-\frac{(n-2)}{B\,z^2\,\ln\,z},&\nu=-1,\\
\frac{n-1}{B\,z},&\nu=-1/2,\\
-\frac{n\,\ln\,z}{B},&\nu=0,\\
\frac{1}{B}\left(1+\frac{n}{2\,\nu}\right),&\nu>0.
\label{gammalims}
\end{array}\right.
\ee
One can see that for $\nu\le0$ the ratio $\gamma_{\nu}(z)$ 
diverges at the origin.
The rate of divergence increases monotonously from
$\ln\,z$ to $z^{-2}$ with the absolute value of $\nu$.

\section{Properties of $K_{\nu}$}
\label{Bessel}
Here we summarize some essential facts about the modified Bessel 
function of the second kind $K_{\nu}(z)$.
The sources of this as well as further information about 
$K_{\nu}(z)$ are handbooks \cite{AS}, Chapters 9, 10 and \cite{GR},
Sect.~8.4.
The function $K_{\nu}$ can be defined as a decreasing solution of 
the modified Bessel differential equation
\[x^2\,y''+x\,y'-\left(x^2+\nu^2\right)\,y=0.\] 
 
The function $K_{\nu}(z)$ exponentially decreases as $z\to\infty$
and diverges at $z=0$. 
In addition, $K_{-\nu}(z)=K_{\nu}(z)$ and 
\be
K_{\nu+1}(z)=K_{\nu-1}(z)+\frac{2\,\nu}{z}\,K_{\nu}(z).
\label{K4}
\ee
For integer $k\ge 0$ we have 
\be
K_{k+1/2}(z) = \sqrt{\frac{\pi}{2z}}e^{-z}\sum_{m=0}^k
\frac{(k+m)!}{m!(k-m)!(2z)^m},
\label{K1}
\ee
and in particular
\be
K_{\pm 1/2}(z)=\sqrt{\frac{\pi}{2\,z}}\,e^{-z};\quad
K_{3/2}(z)=\sqrt{\frac{\pi}{2\,z^3}}\,e^{-z}.
\label{K2}
\ee
For arbitrary fixed $\nu$ and $z\gg\nu$
\be
K_{\nu}(z)\sim\sqrt{\frac{\pi}{2\,z}}\,e^{-z},\quad z\to\infty.
\label{K3}
\ee
The asymptotic behavior at $z=0$ is given by
\be
K_{\nu}(z)\sim\left\{\begin{array}{cc}
\displaystyle\frac{\Gamma(|\nu|)}{2}\,\left(\frac{2}{z}\right)^{|\nu|},&|\nu|\ne0,\\
\displaystyle\log\left(\frac{2}{z}\right)-\gamma,&\nu=0,
\end{array}\right.
\label{lim0}
\ee
where $\gamma\approx 0.577$ is the Euler-Mascheroni constant.

\newpage


%

\begin{thebibliography}{99}

\bibitem{AJK05}
S. Albeverio, V. Jentsch, and H. Kantz (eds),
\emph{Extreme Events in Nature and Society}
(Springer, Heidelberg, 2005).

\bibitem{KBS03}
V. I. Keilis-Borok and A. A. Soloviev, A. A. (eds), 
{\it Nonlinear Dynamics of the Lithosphere and Earthquake Prediction} 
(Springer, Heidelberg, 2003). 

\bibitem{Sor04}
D. Sornette, 
{\it Critical Phenomena in Natural Sciences}
2-nd ed. (Springer-Verlag, Heidelberg, 2004).

\bibitem{Wie49}
N. Wiener, 
{\it Extrapolation, interpolation and smoothing of stationary time series 
with engineering applications},  (Wiley, 1949).

\bibitem{Kol41}
A. N. Kolmogorov, 
{\it Interpolation, extrapolation of stationary random sequences}, 
Izv. Akad. Nauk. SSSR, Ser. Mat., {\it 5}, 3-14 (1941). 
(English translation: W. Doyle, RAND corporation memorandum RM-3090-PR, 
April 1962).

\bibitem{KB61}
R. E. Kalman and R. S. Bucy ,
ASME Transactions, J. Basic Eng., Series D, {\bf 83}, 95-108 (1961).

\bibitem{Kush62}
H. J. Kushner, SIAM J. Control, {\bf 2}, 106-119 (1962). 

\bibitem{Zak69}
M. Zakai, Warsch. Und Ver. Gebiete, {\bf 11}, 230-243 (1969).

\bibitem{DFG01}
A. Doucet, N. de Freitas and N. Gordon (eds),
{\it Sequential Monte Carlo Methods in Practice}, 
(Springer, 2001).

\bibitem{Chow07}
P. L. Chow,
{\it Stochastic Partial Differential Equations}, 
(Chapman Hall/CRC Press, Boca Raton, FL, 2007). 

\bibitem{EKM08}
P. Embrechts, C. Kl\"uppelberg, and T. Mikosch,
{\it Modelling Extremal Events for Insurance and Finance 
(Stochastic Modelling and Applied Probability)}, 
(Springer, 2008).

\bibitem{KKR80}
V. I. Keilis-Borok, L. Knopoff, and I. M. Rotwain, 
Nature {\bf 283}, 258-263 (1980).

\bibitem{KB94}
V. I. Keilis-Borok, 
Physica D {\bf 77}, 193-199 (1994).

\bibitem{KB96}
V. I. Keilis-Borok, 
Proc. Natl. Ac. Sci. USA, {\bf 93}, 3748-3755 (1996).

\bibitem{KB02}
V. I. Keilis-Borok,
Ann. Rev. Earth Planet. Sci., {\bf 30}, 1-33 (2002).

\bibitem{KK90}
V. I. Keilis-Borok and V. G. Kossobokov,
Phys. Earth Planet. Inter. {\bf 61} (1-2), 73-83 (1990).

\bibitem{KS99}
V. I. Keilis-Borok and P. N. Shebalin (eds.), 
Phys. Earth Planet. Inter. {\bf 111}, (1999).

\bibitem{Vor99}
I. A. Vorobieva, 
Phys. Earth Planet. Inter. {\bf 111}, 197-206 (1999).

\bibitem{Aki85}
K. Aki,
Earthquake Prediction Res. {\bf 3}, 219-230 (1985). 

\bibitem{ALP82}
C. J. Allegre, J. L. Lemouel, and A. Provost,
Nature {\bf 297}, 5861, 47-49 (1982).

\bibitem{PA95}
A. Press and C. Allen, 
J. Geophys. Res. {\bf 100}, 6421-6430 (1995).

\bibitem{Rom93}
B. Romanowicz, 
Science {\bf 260}, 1923-1926 (1993).

\bibitem{Mog81}
K. Mogi, 
in {\it Earthquake Prediction: An International Review}, 
Maurice Ewing Series, {\bf 4} (American Geophysical
Union, Washington, DC, 1981), 43-51.

\bibitem{Hab81}
R. E. Haberman, 
in {\it Earthquake Prediction: An International Review}, Maurice Ewing Series, 
{\bf 4} (American Geophysical Union, Washington, DC, 1981), 29-42.

\bibitem{Smi81}
W. Smith, 
Nature {\bf 289}, 136-139 (1981).

\bibitem{YK92}
T. Yamashita and L. Knopoff, 
J. Geophys. Res. {\bf 97}, 19873-19879 (1992).

\bibitem{RTK00}
J. Rundle, D. Turcotte, and W. Klein (eds), 
{\it Geocomplexity and the Physics of Earthquakes.} 
(American Geophysical Union, Washington DC, 2000).

\bibitem{PCS94}
G. F. Pepke, J. R. Carlson, and B. E. Shaw,  
J. Geophys. Res. {\bf 99}, 6769-6788 (1994).

\bibitem{SSS99}
L. R. Sykes,  B. E. Shaw, and  C. H. Scholz,
Pure Appl. Geophys. {\bf 155}:2-4, 207-232 (1999).

\bibitem{Tur91}
D. L. Turcotte,
Ann. Rev. Earth Planet. Sci., {\bf 19}, 263-281 (1991).

\bibitem{JS99}
S. C. Jaume and L. R. Sykes,
Pure Appl. Geophys., {\bf 155} (2-4): 
279-305 (1999).

\bibitem{Jor06}
T. H. Jordan,
Seismol. Res. Lett. {\bf 77}(1), 3-6 (2006).

\bibitem{Lock93}
D. Lockner,
Intl. J. Rock Mech. Mining Sci. Geomech. Abstr. {\bf 30}, 7, 883-899 (1993).

\bibitem{MDR90}
G. Molchan, O. Dmitrieva, and I. Rotwain,
Phys. Earth and Planet. Inter. {\bf 61}, 1-2, 128-139 (1990).

%
\bibitem{ZHK01}
G. Zoeller, S. Hainzl, and J. Kurths, 
J. Geophys. Res. {\bf 106}, 2167–2176 (2001).

\bibitem{EB97b}
M. Eneva and Y. Ben-Zion,
J. Geophys. Res. {\bf 102}, 17785-17795 (1997).

\bibitem{ABR04}
M. Anghel, Y. Ben-Zion, and R. Rico-Martinez,
Pure Appl. Geophys. {\bf 161}, 9-10, 2023-2051 (2004).


\bibitem{RKB97}
I. M. Rotwain, V. I. Keilis-Borok, and L. Botvina, 
Phys. Earth Planet. Inter., {\bf 101}, 61-71 (1997).

\bibitem{KSSM00}
V. Keilis-Borok, J. H. Stock, A. Soloviev, and P. Mikhalev, 
J. Forecasting {\bf 19}, 65–80 (2000).

\bibitem{KSA05}
V. I. Keilis-Borok, A. A. Soloviev, C. B. Allegre CB, {\it et al.} 
Pattern Recognition {\bf 38}, (3), 423-435 (2005). 

\bibitem{ZWG04}
I. Zaliapin, H. Wong, and A. Gabrielov,
Phys. Rev. E, {\bf 71}, 066118 (2004).

\bibitem{GKSZ08}
A. Gabrielov, V. Keilis-Borok, Y. Sinai, and I. Zaliapin,
in \emph{ESI Lecture Notes in Mathematics and Physics: 
Boltzmann's Legacy}, G. Gallavotti, W. Reiter and J. Yngvason (Eds.), 
203-216.

\bibitem{BSL97}
E. M. Blanter,  M. G. Shnirman, and J. L. LeMouel,
Phys. Earth Planet. Inter., {\bf 103} (1-2), 135-150 (1997). 

\bibitem{GZKN00}
A. M. Gabrielov, I. V. Zaliapin, V. I. Keilis-Borok, 
and W. I. Newman W. I., Geophys. J. Int., {\bf 143}, 
427-437 (2000).
 
\bibitem{NS90}
G. S. Narkunskaya and M. G. Shnirman,  
Phys. Earth. Planet. Inter., {\bf 61}, 29-35 (1990).

\bibitem{NS94}
G. S. Narkunskaya and M. G. Shnirman, 
Computational Seismology and Geodynamics, 
(AGU, Washington, D.C.), {\bf 1}, 20-24 (1994).

\bibitem{NTG95}
W. I. Newman, D. L. Turcotte, and A. M. Gabrielov, 
Phys. Rev. E, {\bf 52}, 4827-4835 (1995).

\bibitem{ZKG03}
I. Zaliapin, V. Keilis-Borok, and M. Ghil, 
J. Stat. Phys., {\bf 111}, (3-4), 839-861 (2003). 

\bibitem{Kol41a}
A. N. Kolmogorov, 
Dokl. Akad. Nauk. USSR, {\bf 30}, 299--303, (1941)  (in Russian).
English translation: Proc. Royal Soc. London, Series A, 
{\bf 434}, 9–13 (1991). 

\bibitem{Kol41b}
A. N. Kolmogorov, 
Dokl. Akad. Nauk. USSR, {\bf 32}, 16--18, (1941)  (in Russian).
English translation: Proc. Royal Soc. London, Series A, 
{\bf 434}, 15–17 (1991). 

\bibitem{Obu41}
A. M. Obukhov,
Dokl. Akad. Nauk. USSR, {\bf 1}, 22--24 (1941).

\bibitem{Frisch}
U. Frisch,
{\emph Turbulence: The Legacy of A. M. Kolmogorov},
(Cambridge University Press, 1996).

\bibitem{JCM90}
J. C. McWilliams,
J. Fluid Mech., {\bf 219}, 361--385.

\bibitem{GR41}
B. Gutenberg,  and C. F. Richter,
Geol.\ Soc.\ Amer., Special papers
{\bf 34}, 1-131 (1941).

\bibitem{GR44}
B. Gutenberg, and C.\ F.\ Richter,
Bull.\ Seism.\ Soc.\ Am. {\bf 34} 185-188 (1944).

\bibitem{BZ03}
Y. Ben-Zion, in
\emph{International Handbook of Earthquake and Engineering
Seismology}, Part B, 1857-1875, (Academic Press, 2003).

\bibitem{Zipf}
G. K. Zipf, \emph{Psycho-Biology of Languages}, 
(Houghton-Mifflin, 1935; MIT Press, 1965). 

\bibitem{Pareto}
V. Pareto, 
\emph{Cours d'economie Politique}, (F. Rouge, Lausanne, 1897).

\bibitem{KBL+06}
O. S. Klass, O. Biham, M. Levy, O. Malcai, and S. Soloman, 
Econ. Lett. {\bf 90}, 2, 290–295 (2006). 

\bibitem{Rich}
L. F. Richardson, \emph{Statistics of deadly quarrels},
(Boxwood Pr., 1960).

\bibitem{Lotka}
A. J. Lotka,
J. Washington Ac. Sci. {\bf 16} (12), 317-324 (1926).

\bibitem{MTG+04}
B. D. Malamud, D. L. Turcotte, F. Guzzetti, and P. Reichenbach,
Earth Planet. Sci. Lett. {\bf 229}, 1-2, 45-59 (2004).

\bibitem{BGR09}
M. T. Brunetti, F. Guzzetti, and M. Rossi
Nonlin. Proc. Geophys., {\bf 16}, 2, 179-188 (2009).

\bibitem{BT67}
B. Mandelbrot and H. M. Taylor,
Operation Res., {\bf 15}, 6, 1057-1062 (1967).

\bibitem{PS08}
V. Plerou, H. E. Stanley,
Phys. Rev. E, {\bf 77}, 3, Art. No. 037101 (2008).    

\bibitem{GGP+03}
X. Gabaix, P. Gopikrishnan, V. Plerou, and H. E. Stanley,
Nature, {\bf 423}, (6937), 267-270 (2003).

\bibitem{Bur90}
B. Burlando, J. Theor. Biol. {\bf 146}, 99-114 (1990).

\bibitem{New97}
M. E. J. Newman, Physica D, {\bf 107}, 293-196 (1997).

\bibitem{New05}
M. E. J. Newman,
Contemp. Phys., {\bf 46} (5), 323-351 (2005).    

\bibitem{AB02}
R. Albert and A. L. Barabasi,
Rev. Modern Phys., {\bf 74} (1), 47-97 (2002).

\bibitem{Mandelbrot} 
B. Mandelbrot, 
\emph{The Fractal Geomery of Nature}
(W. H. Freeman, 1983).

\bibitem{Turcotte}
D. L. Turcotte,
\emph{Fractals and Chaos in Geology and Geophysics}
(Cambridge University Press, 2nd ed. 1997).

\bibitem{BTW88}
P. Bak, C. Tang, and K. Wiesenfeld,
Phys. Rev. A, {\bf 38}, 1, 364-374 (1988).

\bibitem{Tur99}
D. L. Turcotte,
Rep. Prog. Phys., {\bf 62}, 10,1377-1429 (1999). 

\bibitem{Dhar90}
D. Dhar,
Phys. Rev. Lett, {\bf 64}, 14, 1613-1616 (1990). 

\bibitem{DS92}
B. Drossel, F. Schwabl,
Phys. Rev. Lett., {\bf 69}, 11, 1629-1632 (1992). 

\bibitem{RK93}
J. B. Rundle and W. Klein,
J. Stat. Phys., {\bf 72}, 1-2, 405-412 (1993). 

\bibitem{BK67}
R. Burridge and L. Knopoff,
Bull. Seism. Soc. Am., {\bf 57}, 3, 341-371 (1967). 

\bibitem{OFC92}
Z. Olami, H. J. S. Feder, and K. Christensen,
Phys. Rev. Lett., {\bf 68}, 8, 1244-1247 (1992) 

\bibitem{AN04}
K. B. Athreya and P. E. Ney,
\emph{Branching Processes}
(Dover Publications, 2004).

\bibitem{Eva98}
L.C. Evans, 
\emph{Partial Differential Equations}
(American Mathematical Society, Providence, 1998).

\bibitem{Stanley71}
H. E. Stanley,
\emph{Introduction to Phase Transitions and Critical Phenomena}
(Oxford University Press, 1971). 

\bibitem{Ma00}
S.-K. Ma,
\emph{Modern Theory of Critical Phenomena}
(Westview Press, 2000)

\bibitem{Kadanoff00}
L. P. Kadanoff,
\emph{Statistical Physics: Statics, Dynamics and Renormalization}
(World Scientific Publishing Company, 2000).

\bibitem{SA94}
D. Stauffer and A. Aharony,
\emph{Introduction to Percolation Theory}
(CRC, 1994).

\bibitem{Grimmett}
G. R. Grimmett,
\emph{Percolation}
(Springer, 2nd ed, 1999).

\bibitem{Bollobas}
B. Bollob\'as,
\emph{Random Graphs}
(Cambridge University Press, 2nd ed, 2001).

\bibitem{Durrett}
R. Durrett,
\emph{Random Graph Dynamics}
(Cambridge University Press, 2006).

\bibitem{NBW06}
M. E. J. Newman, A.-L. Barabasi, and D. J. Watts,
\emph{The Structure and Dynamics of Networks}
(Princeton University Press, 2006).

\bibitem{AS}
Abramowitz, M. and Stegun, I. A. (Eds.),
{\it Handbook of Mathematical Functions with Formulas, 
Graphs, and Mathematical Tables} (New York: Dover, 1965). 

\bibitem{GR}
I. S. Gradshtein and I. M. Ryzhik,
{\it Tables of Integrals, Series and Products.}
(A. Jeffrey and D. Zwillinger (eds.) Academic Press, 7-th ed., 2007)
 
\end{thebibliography}
\end{document}